\documentclass[prb,aps,twocolumn,amsmath]{revtex4-1}
\usepackage{graphicx}
\usepackage{color}
\usepackage{multirow}
\usepackage{hhline}
\usepackage{soul}
\usepackage{bm}

\definecolor{orange}{rgb}{1,0.5,0}
\usepackage{siunitx}

\begin{document}

\title{Numerical modelling of the coupling efficiency of single quantum emitters in photonic-crystal waveguides}
\author{Alisa Javadi}\email{javadi@nbi.ku.dk} \homepage{www.quantum-photonics.dk}
\author{Sahand Mahmoodian}
\author{Immo S\"{o}llner}
\author{Peter Lodahl}

\affiliation{Niels Bohr Institute, University of Copenhagen, Blegdamsvej 17, DK-2100 Copenhagen, Denmark}

\date{\today}

\pacs{42.50.-p, 78.67.Hc, 78.47.-p, 42.50.Ct, 78.67.Pt}

\begin{abstract}
Planar photonic nanostructures have recently attracted a great deal of attention for quantum optics applications. In this article, we carry out full 3D numerical simulations to fully account for all radiation channels and thereby quantify the coupling efficiency of a quantum emitter embedded in a photonic-crystal waveguide. We utilize mixed boundary conditions by combining active Dirichlet boundary conditions for the guided mode and perfectly-matched layers for the radiation modes. In this way, the leakage from the quantum emitter to the surrounding environment can be determined and the spectral and spatial dependence of the coupling to the radiation modes can be quantified. The spatial maps of the coupling efficiency, the $\beta$-factor, reveal that even for moderately slow light, near-unity $\beta$ is achievable that is remarkably robust to the position of the emitter in the waveguide. Our results show that photonic-crystal waveguides constitute a suitable platform to achieve deterministic interfacing of a single photon and a single quantum emitter, which has a range of applications for photonic quantum technology. \end{abstract}

\maketitle

\section{Introduction}

Enhancing the spontaneous emission rate of a quantum emitter by placing it in an optical cavity was first suggested by Purcell \cite{Purcell1946PR}. In the following decades it was realized that the spontaneous emission rate of a quantum emitter can also be suppressed by placing it in a photonic bandgap \cite{Bykov1975, Yablonovitch1987PRL,Lodahl2004Nature}. This has led to a significant research effort in manipulating the photonic environment surrounding quantum emitters to suppress coupling to unwanted radiation modes and to boost coupling to specific localized modes. The spontaneous emission rate of a quantum emitter scales with the projected local density of optical states (LDOS). Significant enhancement of spontaneous emission rates have been demonstrated in optical cavities \cite{Gerard1998PRL}, nanophotonic waveguides \cite{Lund-Hansen2008PRL} and with surface plasmon modes \cite{Akimov2007Nature}, while suppression of spontaneous emission has been measured in the bandgap region of a photonic crystal \cite{Wang2011PRL}.

Recently there has been a growing interest in quantum emitters coupled to planar nanostructures. Indeed different quantum emitters such as quantum dots \cite{Lodahl2015RMP,Javadi2015NCOM,Pinotsi2001IEEEJQP,Carter2013NPHOT,makhonin2014waveguide}, diamond color centers \cite{Loncar2013MRS,Sipahigil2016Science}, {and atoms \cite{Tiecke2014Nature,Goban2014Ncom}} have been efficiently coupled to planar nanostructures. Planar photonic crystals typically only possess a bandgap for a single polarization and for in-plane guided propagation. Nevertheless, this partial bandgap can greatly reduce the LDOS for emitters oriented in plane by suppressing the coupling rate to the radiation modes and therefore decrease the spontaneous emission rate of embedded quantum emitters \cite{Koenderink2006JOSAB,Wang2011PRL}. By implementing  waveguides or cavities in the band gap frequency region, the spontaneous emission can preferentially be directed with very high efficiency into a single mode. Combination of suppression of the coupling to the radiation modes and the enhancement of coupling to a photonic-crystal waveguide (PCW) mode has been predicted to enable a deterministic single-photon source \cite{MangaRao2007PRL, LeCamp2007PRL}. The fraction of emitted light that is coupled into the waveguide  is defined as the $\beta$-factor. The coupling of single quantum dots to PCWs has been studied by several groups \cite{Lund-Hansen2008PRL,Dewhurst2010APL,Hoang2012APL,Laucht2012PRX} and a record value of $\beta > 98.4\%$ was recently achieved \cite{Arcari2014PRL}.

{ An important feature of PCWs is their wide bandwidth contrary to cavities. However, the $\beta$-factor depends significantly on the spatial position of the emitter in the PCW due to the coupling to waveguide mode as well as the coupling to radiation modes.} The spatial and spectral dependencies of the coupling to the PCW guided mode  are well understood \cite{JoannopoulosBook} since they can be obtained from eigenfunctions computed using standard techniques, e.g.,the plane-wave expansion method \cite{JoannopoulosBook,Johnson2001OE}. In contrast, the spatial and spectral dependencies of the radiation continuum in PCWs have thus far only been quantified at certain spatial positions \cite{MangaRao2007PRB,LeCamp2007PRL}. A full mapping of the radiation modes is essential in order to find the $\beta$-factor and thereby determine how large coupling efficiencies may be obtained under experimentally realistic conditions.

 In this article, we develop the necessary tools to carry out a detailed analysis of the LDOS in a PCW. The main challenge in modeling an infinite PCW in a finite computation domain is that, along the propagation direction, open boundary conditions are required. Although perfectly matched layers are typically good approximations for open boundaries, they fail in inhomogeneous dielectric structures \cite{Oskooi2008OE}, particularly at low group velocities. To overcome this problem, we derive Dirichlet boundary conditions, whose phase and amplitude match a propagating PCW mode excited by a dipole at an arbitrary point inside the waveguide. Armed with these boundary conditions, we compute the LDOS contribution from the radiation continuum for a range of frequencies across the waveguide band. We show that the coupling to the radiation continuum is highly suppressed in a PCW. We map out the dependence of the coupling to the radiation continuum on the position, frequency, and orientation of the dipole. The resulting $\beta$-factor is remarkably robust to spatial position and spectral tuning of the emitter, which has been confirmed experimentally \cite{Arcari2014PRL}.

This paper is arranged in the following sections: Section \ref{electrodynm} discusses the different decay channels for an emitter embedded in a PCW and introduces the parameters that govern the emitter dynamics. Section \ref{radmodes} includes the details of the simulations. We present and discuss the results of the numerical simulations in Section \ref{results}. Section \ref{conclusions} sums up our results and gives an overview of various applications that could benefit from an efficient light matter-interface. The two appendices include the convergence tests, and a short overview of the decay dynamics of an emitter in a photonic crystal and the comparison to a PCW.

\section{Electrodynamics of a quantum emitter in a PCW\label{electrodynm}}

\begin{figure}
  \includegraphics[width=86 mm]{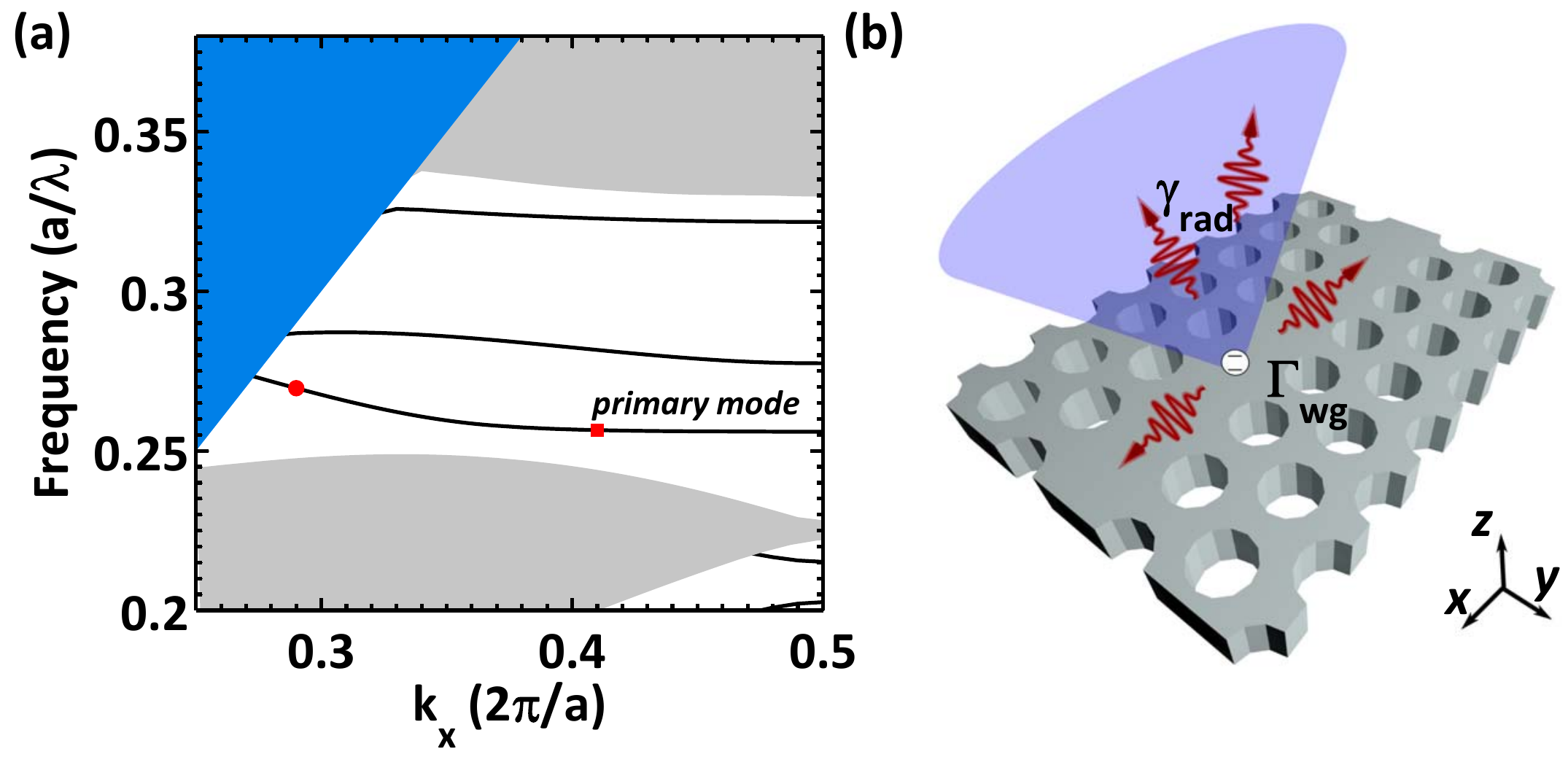}\\
  \caption{(a) Band diagram of a PCW in a membrane for transverse electric modes (TE). The solid black lines are the guided modes of the waveguide. The gray regions mark the membrane guided modes. The blue region is the continuum of the radiation modes that are not bounded to the membrane. The red circle and square mark the frequencies corresponding to $n_g=5$ and $n_g=58$ respectively. (b) Sketch of a quantum emitter in the middle of a PCW showing the coupling to radiation continuum ($\gamma_\textrm{rad}$) and to the guided mode ($\gamma_\textrm{wg})$.}\label{band_struct}
\end{figure}

Figure \ref{band_struct}a, shows the band diagram of the TE modes of a PCW membrane\cite{JoannopoulosBook}. Inside the band gap, light is mainly guided by three highly confined waveguide modes and by matching the PCW to the targeted emitter, the emitter is typically coupled to a single propagating mode (cf. Fig. \ref{band_struct}a, solid black lines). The waveguide modes are highly dispersive and the group velocity of the wave is reduced as its frequency approaches the band edge, where the slow-down factor $n_g$ (also known as the group index) ideally diverges. Due to the partial band gap of the 2D PCW membrane, there exist a continuum of modes that are not guided by the waveguide and leak to the surrounding environment (blue area in Fig. \ref{band_struct}a). In real PCWs unavoidable fabrication imperfections influence light transport leading to multiple scattering effects. As a consequence, the guided mode is coupled to leaky modes or back-scattered to the oppositely propagating mode in the waveguide \cite{Hughes2005PRL,Mazoyer2009PRL,savona2011PRB}, which has been quantitatively studied in Ref. \cite{Smolka2011NJP}. Effects of disorder become dominating for long waveguides and large group indices, and may be eliminated by reducing both. In the present work we only consider PCWs where effects of disorder are negligible, which in practice means that we consider quantum emitters coupled only to moderately large values of $n_g$ (in experiments typically  $n_g \sim 100$ can be achieved).

An emitter embedded in a PCW can emit photons either to the guided modes of the PCW or to the radiation continuum, as schematically illustrated in Fig. \ref{band_struct}b. The spontaneous emission rate of an excited emitter can be related to the transition dipole moment $\boldsymbol{d}$, and the projected LDOS as \cite{NanoOpticsBook,Lodahl2015RMP}:
\begin{equation}
\label{eq:decay_rate}
\begin{aligned}
& \gamma=\frac{\pi\omega}{\hbar\epsilon_{0}}|\boldsymbol{d}|^2\rho(\omega_0,\mathbf r_0,\mathbf{n_d}),
\end{aligned}
\end{equation}
where
\begin{equation}
\label{eq:projected_LDOS}
\begin{aligned}
& \rho(\omega_0,\mathbf r_0,\mathbf{n_d})=\sum\limits_\mathbf{k}|\mathbf{n_{d}}\cdot \mathbf{u^*_{k}}(\mathbf r_0)|^2\delta(\omega_0-\omega_\mathbf{k}),
\end{aligned}
\end{equation}
where $\mathbf{n_d}$ is the orientation of the the dipole moment, $\rho(\omega_0,\mathbf r_0,\mathbf{n_d})$ is the projected LDOS, and $\mathbf{u_{k}}$ denotes the electric field eigenmode functions.

The modes in Fig. \ref{band_struct}a are classified into three categories: the guided modes of the PCW, the slab guided modes present outside the bandgap region (the gray region in Fig. \ref{band_struct}b), and the continuum of radiation modes. At each frequency, the total decay rate of the emitter can be written as a sum of the contribution from these three sets of modes in addition to any residual contributions from coupling to transverse magnetic (TM) modes, i.e. $\gamma_\textrm{total}(\omega)=\gamma_\textrm{wg}(\omega)+ \gamma_\textrm{rad}(\omega)+\gamma_\textrm{slab}(\omega)+\gamma_\textrm{TM}(\omega),$
where the explicit dependence on spatial position and dipole orientation has been omitted for brevity.
The $\beta$-factor quantifying the fraction of radiation coupled to the primary waveguide mode is defined as
$\beta=\frac{\gamma_\textrm{wg}}{\gamma_\textrm{total}}.$
For the purpose of this paper, we limit the discussion to the experimentally relevant situation of dipole emitters located in the center of the membrane, whereby no coupling to TM modes is present, i.e. $\gamma_\textrm{TM} = 0$. The frequency range of interest for a PCW single-photon source is mainly the primary guided mode in the bandgap region, where also $\gamma_\textrm{slab}=0$.

 The contribution to the $\beta$-factor describing the coupling to the waveguide, $\gamma_\textrm{wg}$, is straightforwardly determined by computing the eigenvalue and the corresponding eigenvectors of the electric and magnetic fields $\mathbf{E}_{pg}(\omega,r)$, $\mathbf{H}_{pg}(\omega,r)$ \cite{MangaRao2007PRB,De2012JOSA,LeCamp2007PRL,Chen2010PRB}. The corresponding Purcell factor is defined as
\begin{equation}
\label{eq:gamma_wg}
\begin{aligned}
& F_{p}^\textrm{wg}= \frac{\gamma_\textrm{wg}}{\gamma_0} = \frac{6\pi^2c^3\epsilon_0|\mathbf{E}_\textrm{pg}\cdot\mathbf{n_d}^*|^2}{\omega^2\int\limits_\textrm{unitcell} d^3r {n\operatorname{Re}[ \mathbf{E_\textrm{pg}}\times \mathbf{H_\textrm{pg}}^*}]/a},
\end{aligned}
\end{equation}
where $n$ is the refractive index of the membrane material, $\gamma_0=\frac{n\omega^3 d^2}{3\pi\epsilon_0\hbar c^3}$ is the decay rate of an emitter in a homogenous material of refractive index $n$.

\section{Computing the coupling of a dipole to the radiation continuum\label{radmodes}}
\begin{figure}[ht!]
  \includegraphics[width=86mm]{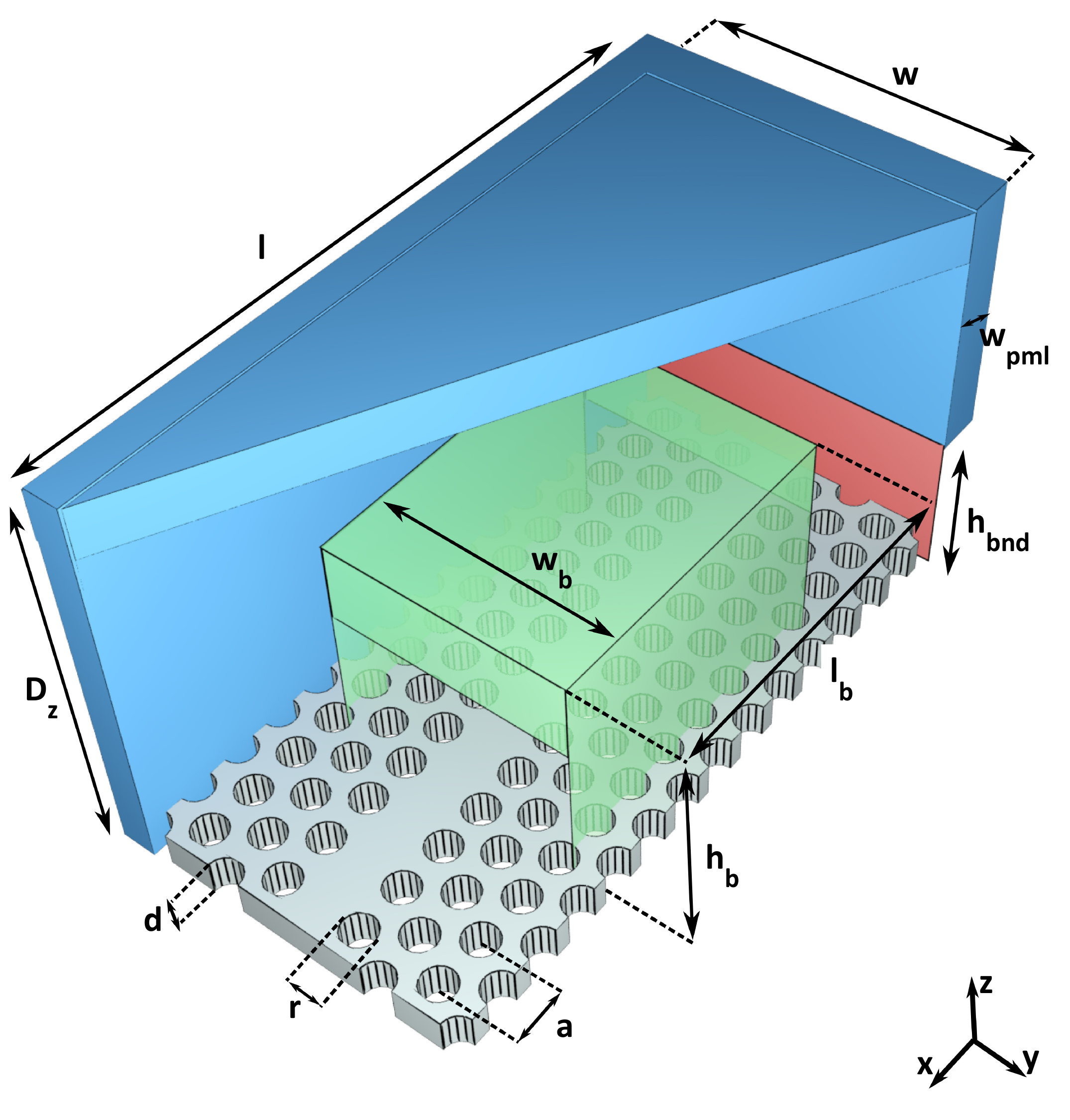}\\
  \caption{A cut through the simulation domain. The blue box is the PML layer around the air domain. The green box is the integration surface that captures the radiation modes and the red plane is where the active boundary conditions are applied. }\label{sim_domain}
\end{figure}
In this section we detail how to extract the contribution to $\beta$ of coupling to the radiation continuum, i.e. $\gamma_\textrm{rad}$, which numerically is the most challenging part of the problem.  The coupling is quantified by the Purcell factor of coupling to the radiation modes, which is denoted $F_p^\mathrm{rad}$. It is given by $F_p^\mathrm{mode}=\gamma^\mathrm{mode}/\gamma_0=P^\mathrm{mode}/P_0$, where $\gamma^\mathrm{mode}$ is the rate of coupling to radiation modes and $P_0$ and $P^\mathrm{mode}$ are the power emitted from the dipole in the reference medium and the nanophotonic structure, respectively.
The total power emitted from the dipole can be extracted by integrating Poynting's vector over a closed surface around the current source, i.e. $P_\textrm{total}=1/2\mathrm{Re} \left[\oint d\mathbf s \mathbf{E} \times \mathbf{H}^* \right]$. Thereby the Purcell factor can be determined.

A main consideration in numerical simulations of optical problems is to ensure proper convergence, i.e., the computed quantities must not depend on the physical size of the computational domain. At the same time, it is desirable to limit the geometrical size of the simulation domain to the minimum possible in order to make the simulation efficient. A general approach to tackle these problems has been to introduce an absorber in the boundaries of a finite simulation domain and adiabatically absorb the incoming wave\cite{berenger1994perfectly,chew19943MOTL}. This can be applied when the geometry of the computational domain is invariant in the direction perpendicular to the boundary and the solutions are propagating waves rather than evanescent fields. In the case of a PCW the simulation domain is invariant along $y$ and $z$ at the boundaries, hence we can apply such perfectly-matched-layer (PML) boundary conditions. However, this is not applicable along the propagation direction ($x$) in the PCW. The generalization of PMLs to photonic- crystal waveguides is challenging\cite{Oskooi2008OE}, particularly for slowly-propagating Bloch modes.

Instead of using PMLs along the direction of the waveguide ($x$), a better choice is to introduce Dirichlet boundary conditions for the purpose of mimicking an open system. This corresponds to setting $E|_{x\pm}=C_\pm$ at the two ends of the waveguide ($x_\pm$). In general, $C_\pm$ have  contributions both from the primary mode of the waveguide and the radiation modes, however the main contribution stems from the guided mode of the waveguides that are extended by many optical wavelengths, i.e. the contributions from radiation modes are negligible. This is checked explicitly  by running a convergence test while varying the length of the simulation domain. The electric fields at the right and left boundaries ($x_\pm$) can be written as
\begin{equation}
\label{eq:A_j_lineardipole}
\begin{aligned}
& E|_{{x}\pm}=-|A^{{r(l)}}_0|e^{-i\phi(\mathbf r_0)} \mathbf{E}^{*}_\textrm{pg}e^{\pm ikx_{\pm}},
\end{aligned}
\end{equation}
where $\mathbf r_0$ is the position of the dipole in the unitcell and $\phi(\mathbf r_0)$ is the phase of the projection of $\mathbf{E_\textrm{pg}}$ on the dipole at the position $\mathbf r_0$. These boundary conditions are referred to as \textit{active boundary conditions}. The amplitudes $A^{r(l)}_0$ can be calculated from the knowledge of the eigenvectors of the PCW using the Green function formalism. For linear dipoles, these amplitudes simplify to:
\begin{equation}
\label{eq:A_j}
\begin{aligned}
& |A^{r}_0|=|A^{l}_0|=\sqrt{\frac{F_p^\textrm{wg}P_0}{1/a \int\limits_\textrm{unitcell}{d^3r  \textrm{Re}[\mathbf{E_\textrm{pg}}^*\times \mathbf{H_\textrm{pg}}]}}},
\\
& \phi=\textrm{arg}(-i \mathbf{E_\textrm{pg}}(\mathbf r_0) \cdot \mathbf{d}).
\end{aligned}
\end{equation}
$\gamma_\textrm{rad}$ can subsequently be calculated as the difference between $\gamma_\textrm{wg}$ and $\gamma_\textrm{total}$. However, for a PCW typically $\gamma_\textrm{rad}\ll \gamma_\textrm{wg}$ and even small reflections, and numerical inaccuracies in $\gamma_\textrm{wg}$ or $\gamma_\textrm{total}$ limit the obtainable precision of $\gamma_\textrm{rad}$. This can be circumvented by calculating $\gamma_\textrm{rad}$ directly by integrating Poynting's vector over a box surrounding the current source and leaving out the integration over the boundaries normal to waveguide direction. This is indicated by the green box in Fig. \ref{sim_domain}, which illustrates the geometry of the computation domain.
Due to the symmetry of the structure and the position of the dipole being in the center of the slab, the solutions of Maxwell's equations are eigenvectors of the mirror symmetry operator about the $z=0$ plane. As a result, the simulation domain can be cut in half along this symmetry plane with the following boundary conditions: $E_{z}(z=0)=0$ and $\frac{\partial }{\partial z}\{E_{x},E_{y}\}|_{z=0}=0$.
 The PCW membrane has a length of $l=(2n+1)a$ and a width of $w=\sqrt{3}(2m+1)a$ and is surrounded by an air box of hight $D_{z}$. The refractive index of the PCW slab is chosen to be $3.5$ corresponding to the refractive index of  GaAs. The simulation domain is encapsulated by PMLs on all sides (blue box in Fig. \ref{sim_domain}). The width of the PML layer is $W_\textrm{PML}$. Active boundary conditions override the PMLs on the two ends of simulation domain normal to the waveguide direction (red plane in Fig. \ref{sim_domain}). The height of these planes are $h_\textrm{bnd}$ and they cover the full waveguide in the $y$ direction. The green box in Fig. \ref{sim_domain} resembles the box that captures the radiation modes. $l_b$, $w_b$ and $h_b$ are the length, width and height of the radiation box. Note that Fig. \ref{sim_domain} is not to scale.

\begin{table}[h]
\caption{Parameter list}
\label{tbl:pram_list}
\begin{tabular}{ccccccccccccccccccc}
\hline
\hline
\multicolumn{1}{|c|}{Parameter} &$l$   &  & $w$         &    & $D_z$       & &$l_b$  &    &$w_b$     &    &$h_b$  &   &$h_{bnd}$    & \multicolumn{1}{c|}{} \\ \hline
\multicolumn{1}{|c|}{Value}     &$33a$ &  &$9\sqrt{3}a$  &   &$6.6a$ & &$31a$ &    &$8\sqrt{3}a$  &   &$2.5a$ & & $2a$     & \multicolumn{1}{c|}{} \\ \hline

\end{tabular}
\end{table}

Table \ref{tbl:pram_list} presents the parameter values that were used in the computations presented in the article. To establish these numbers we have carried out rigorous convergence tests. Appendix \ref{ap:Convergance_tests} contains the results of some of the convergence tests for the most sensitive parameters, $l$, $l_\textrm{bnd}$ and $h_\textrm{bnd}$. From the convergence test results, we estimate that the values of $\gamma_\textrm{rad}$ are accurate to within $5\%$.

The simulation procedure can be summarized as follows: We first carry out an eigenvalue calculation to determine the eigenfrequency, the group index $n_g$, eigenvector of the primary guided mode, and $F_p^\textrm{wg}$ for a given dipole position. Using Eq. (\ref{eq:A_j_lineardipole}) we determine the correct amplitudes for the respective boundaries of the waveguide. Subsequently a finite element frequency domain simulation of a dipole in a PCW is performed with the correct boundary conditions. The total power emitted from the waveguide is calculated by integrating Poynting's vector over a small box around the dipole. The coupling rate, $\gamma_\textrm{rad}$ is extracted by integrating Poynting's vector over the radiation box. We repeat all the simulations for $n_g=5$, $20$, $58$, $120$. These correspond to realistic values of the slow-down factor of light, which have been obtained experimentally in GaAs PCWs\cite{Arcari2014PRL,Wasley2012APL}.

\section{Results\label{results}}
\begin{figure}[h!]
  \includegraphics[width=86mm]{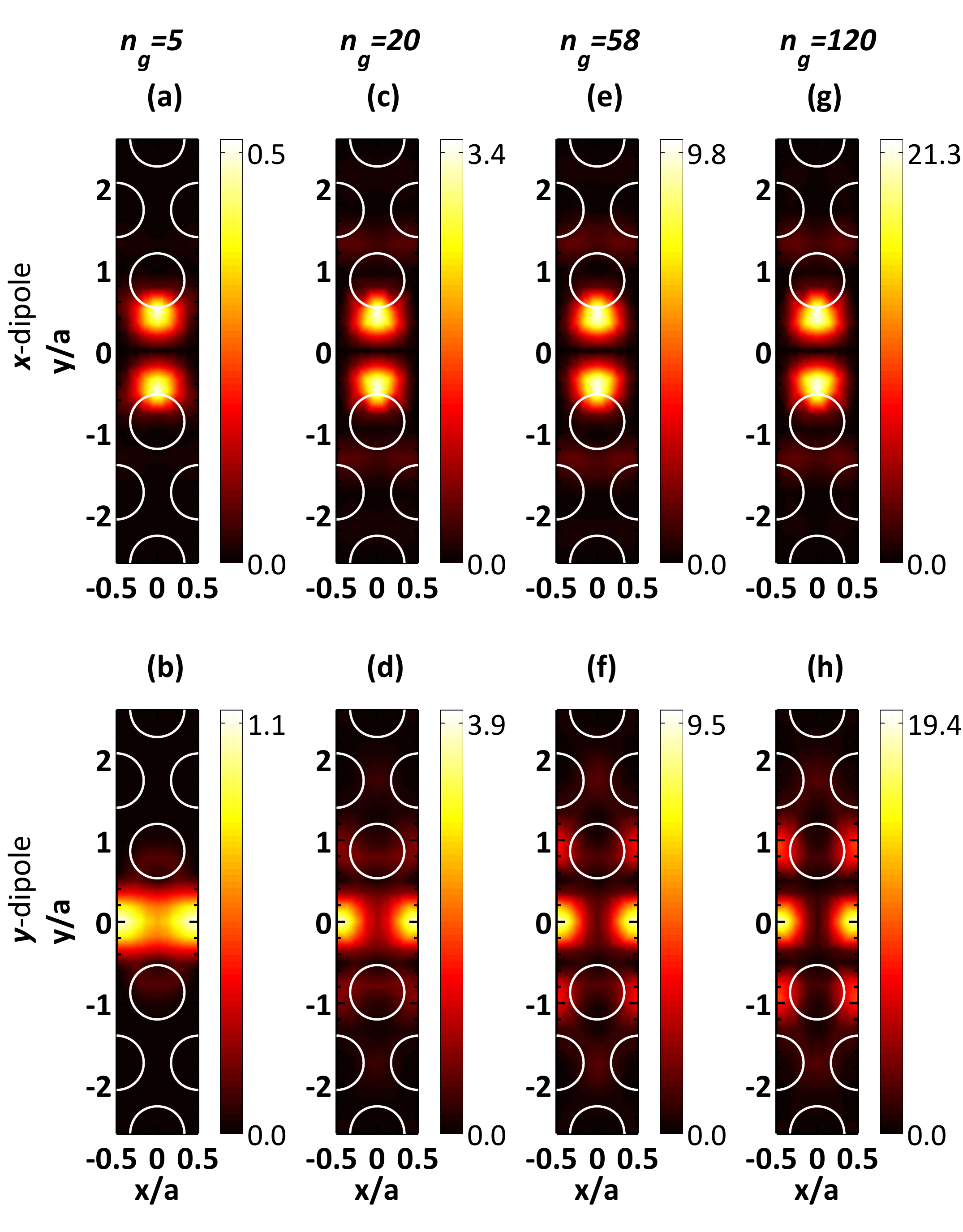}\\
  \caption{ The spatial map of the waveguide Purcell factor, $F_p^\textrm{wg}$, for x/y dipole orientations (upper/lower row) and group indices of $n_g=5$ (a-b), $n_g=20$ (c-d), $n_g=58$ (e-f), and $n_g=120$ (g-h). The white circles represent the air holes. The light-matter interaction is enhanced as the light propagation slows down, and hence the maximum value of $F_p^\textrm{wg}$ increases.  The spatial dependence of $F_p^\textrm{wg}$ follows the Bloch mode of the PCW. }\label{spaitial_map_fpwg}
\end{figure}

 The Purcell factor of a quantum emitter coupled to a waveguide is an important figure-of-merit determining the rate of photon generation and ability to overcome decoherence processes. Figure \ref{spaitial_map_fpwg} shows the position and frequency dependence of $F_p^\textrm{wg}$ for $x$- and $y$-oriented dipoles.  The four columns correspond to dipoles at different frequencies, $n_g=5$, $20, 58,$ and $120$, respectively. At $n_g=5$ the Purcell factor is less than one, but scales linearly with the group index and reaches 23 at $n_g=120$. At the bandedge of the waveguide, the group index and consequently the Purcell-factor diverge. However, in practice this van-Hove singularity in the LDOS is damped by Anderson localization of light induced by unavoidable  fabrication disorder. Experimentally $F_p^\textrm{wg}=24$ has been reported for quantum emitters coupled to PCWs\cite{Javadi2014OE}.

\begin{figure}[ht!]
  \includegraphics[width=86mm]{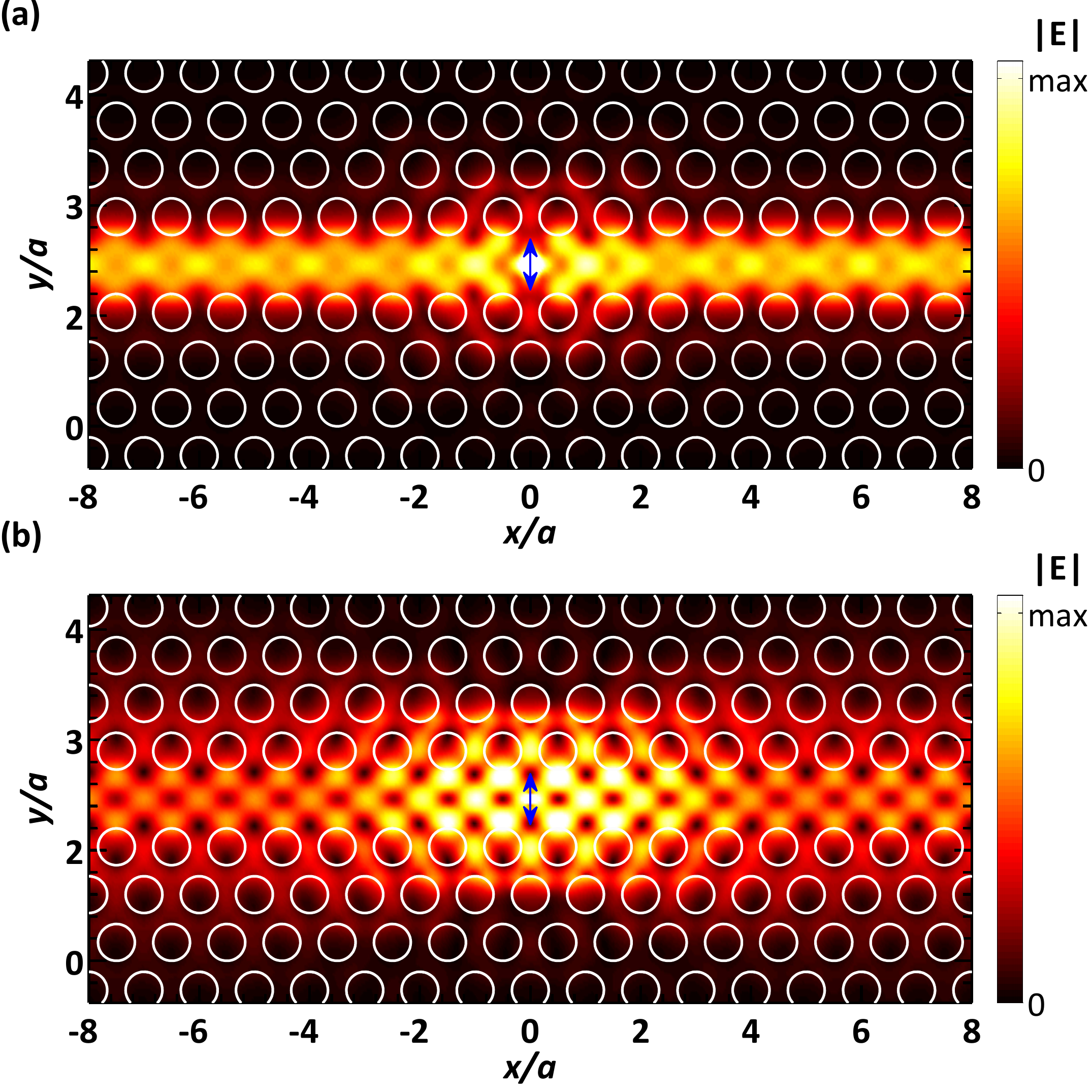}\\
  \caption{(a) Spatial map of the magnitude of the electric field generated by a $y$-dipole placed in the anti-node of $E_y$ and for $n_g=5$ and (b) $n_g=58$. The blue arrow shows the dipole and its orientation. The color scale is saturated at the point of the dipole.}\label{field_profile}
\end{figure}

The actual excitation of the waveguide mode by a dipole emitter is shown in Fig. \ref{field_profile}, which plots $|\textbf{E}|$ for a $y$-oriented dipole in the antinode of the $E_y$ field for $n_g=5$ and $n_g=58$, corresponding to fast and slow light propagation in the PCW. The plots are zoom-ins around the position of the dipole and it should be mentioned that the color bars have been saturated since $|\mathbf{E}|$  diverges at the position of the point source.
Furthermore, a 'chevron feature' in the field profile is observed close to the dipole, which is a manifestation of dipole-induced light localization coming from the coupling to evanescent modes of the PCW  \cite{Bykov1975,John1991PRB,Douglas2015NPHOT,Munro2016Arxiv,Calajo2016PRA}. We observe that the active boundary conditions suppress the reflections from the boundaries of the simulation domain very effectively, i.e. the field intensities on the right hand side and left hand side of the simulation domain are uniform as expected from an infinite system.  The field profiles plotted in Fig. \ref{field_profile}  are subsequently integrated, as detailed in the previous section, in order to extract coupling to radiation modes.

\begin{figure}[h!]
  \includegraphics[width=86mm]{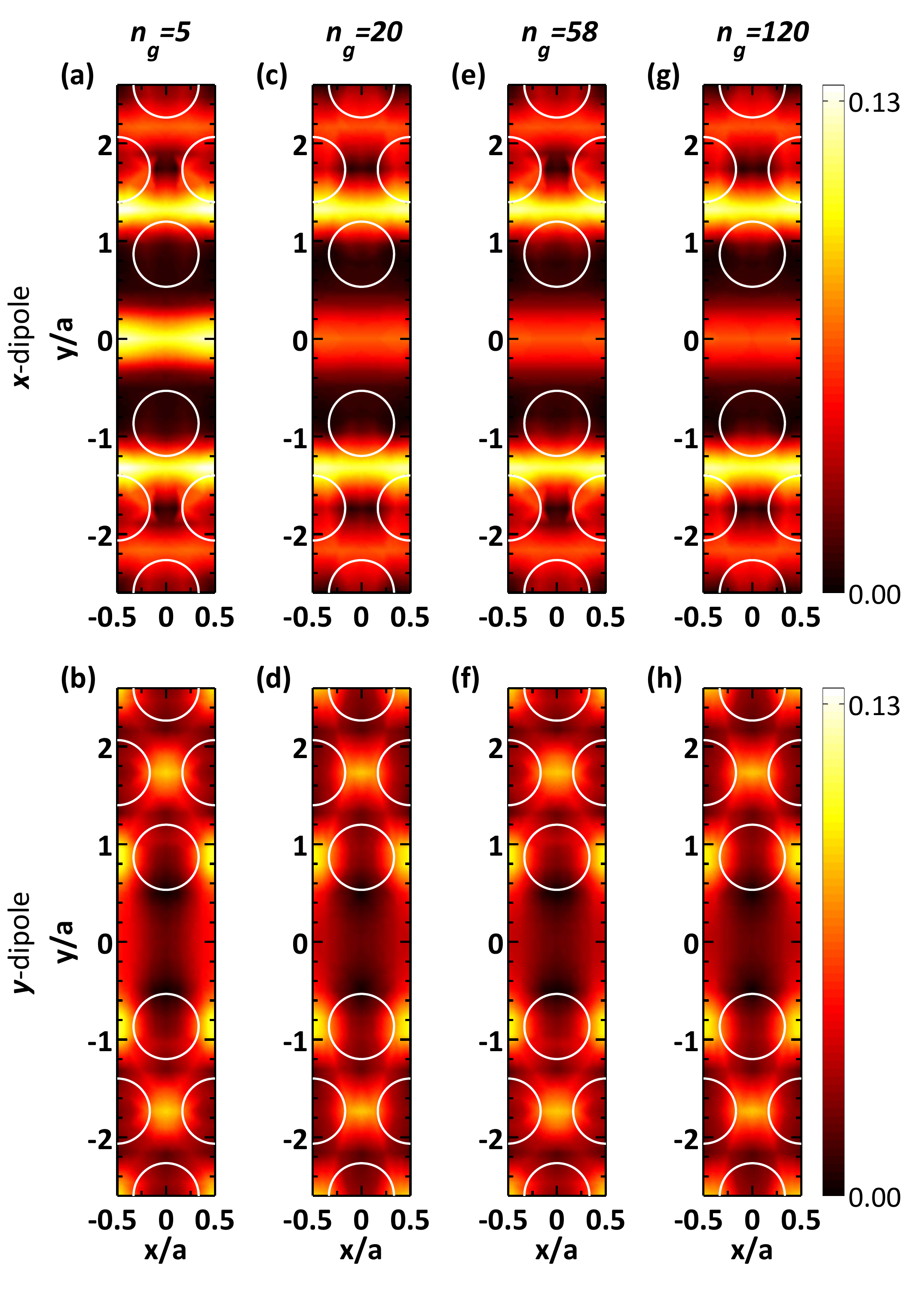}\\
  \caption{ The map of the coupling to the radiation modes as quantified by $F_p^\textrm{rad}$, for $x$- and $y$-dipole orientations. (a-b) at $n_g=5$, (c-d) at $n_g=20$, (e-f) at $n_g=58$, and (g-h) at $n_g=120$. We find $F_p^\textrm{rad} \leq 0.13$ for all positions and a minimum value of $F_p^\textrm{rad} = 0.005.$ }\label{spaitial_map_gammarad}
\end{figure}

\begin{figure}[h!]
  \includegraphics[width=86mm]{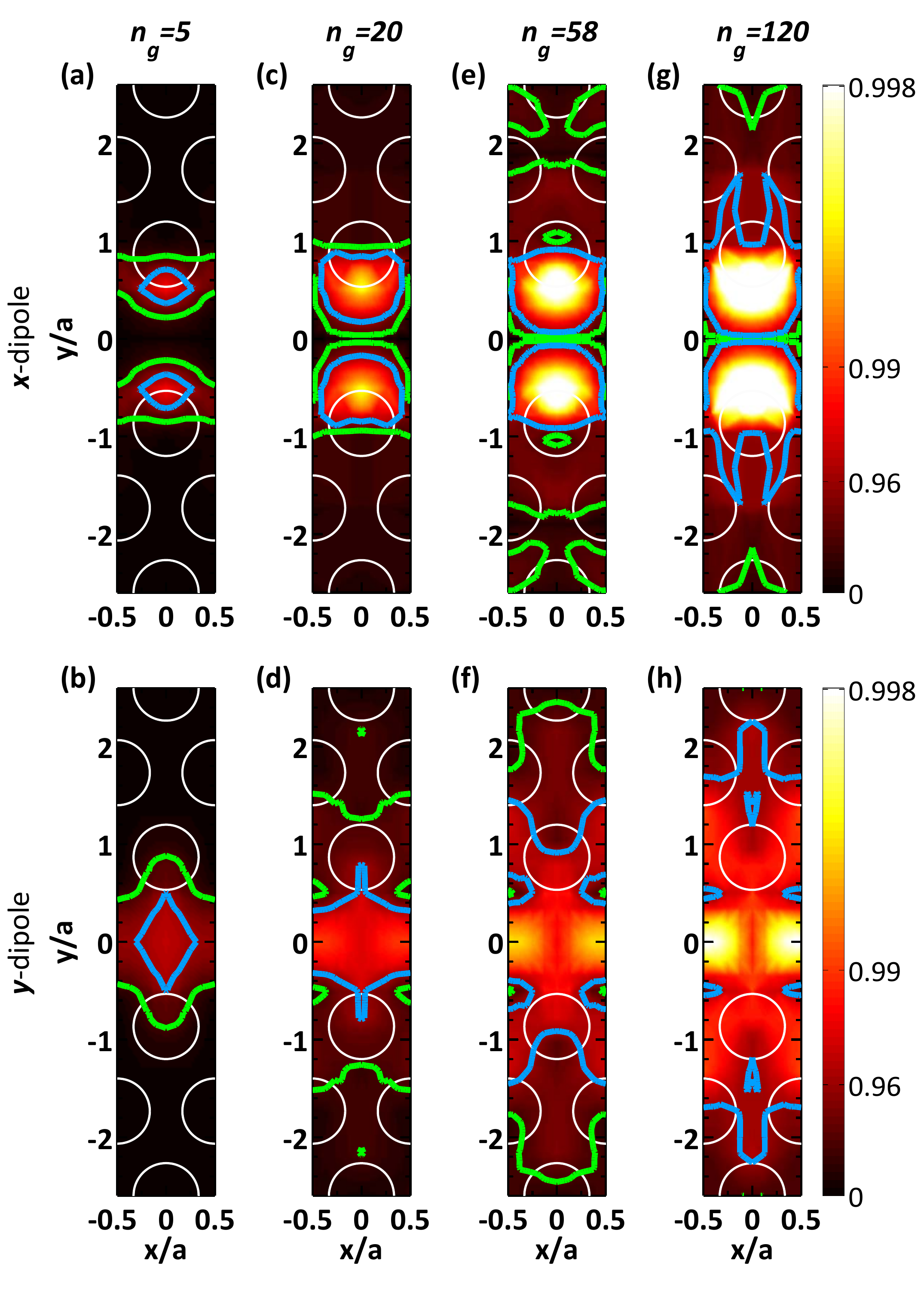}\\
  \caption{ The map of the $\beta$-factor for $x$ and $y$ oriented dipoles. (a-b) at $n_g=5$, (c-d) at $n_g=20$, (e-f) at $n_g=58$, and (g-h) at $n_g=120$. The green and the blue contours correspond to $\beta=0.8$ and $\beta=0.96$, respectively. Note that the highly nonlinear scale bar, i.e. $\beta$ is close to unity in very large spatial ranges. }\label{spaitial_map_beta}
\end{figure}

Subsequently we discuss the results of the computations of the coupling to radiation modes. Figure \ref{spaitial_map_gammarad} shows the position dependence of the Purcell factor associated with the coupling to radiation continuum, $F_p^\textrm{rad}={\gamma_{\textrm{rad}}}/{\gamma_0}$, inside one unit cell. In this case it is desirable to reduce $F_p^\textrm{rad}$ as much as possible below unity, so that the parasitic coupling to non-guided modes is reduced. We find that the suppression is better than a factor of 10 for most spatial positions in Fig. \ref{spaitial_map_gammarad}, and importantly $F_p^\textrm{rad}$ has a complex spatial structure. On the contrary, the frequency dependence of $F_p^\textrm{rad}$ is rather weak and, e.g., changes only about 10\% for a $y$-oriented dipole between $n_g=5$ and $n_g=120$. The smallest achievable Purcell factor is  $F_p^\textrm{rad} = 0.005$, i.e. suppression of radiation modes by a factor of $200$ relative to the emission rate of a homogeneous medium. The strong suppression of radiation modes in 2D photonic-crystal membranes was first predicted in Ref. \cite{Koenderink2006JOSAB} for photonic crystals without defects. Interestingly the suppression achieved in a PCW reaches the value obtainable in a photonic crystal without defects demonstrating that the missing row of holes in the PCW does not induce additional leakage of the light from the membrane, see Appendix \ref{ap:PhC_radiationmodes} for further details.

Finally, the spatial map of the $\beta$-factor and its frequency dependence is investigated, see Fig. \ref{spaitial_map_beta}. Here the green and the blue contours correspond to $\beta=0.80$ and $\beta=0.96$. Note that the implemented color bar showing the magnitude of $\beta$ is highly nonlinear. Even at low $n_g$, cf. Fig. \ref{spaitial_map_beta}(a and b), a large $\beta$-factor can be achieved (higher than $96\%$) although limited to relatively small spatial regions in the PCW. Increasing $n_g$ by moving into the slow-light region, cf. Fig. \ref{spaitial_map_beta}(c-h), increases $\beta$ significantly, and we find $\beta \geq 0.96$ for a very wide range of  dipole positions. More quantitatively, for any dipole located within $\pm a$ from the center of the waveguide $\beta \geq 0.96$ at the experimentally achievable value of $n_g=58$. This is a remarkably robustness towards spatial and spectral detuning, which was already confirmed experimentally where the statistics of the $\beta$-factor of more than 70 different quantum dots in a PCW has been reported \cite{Arcari2014PRL}.

\section{Conclusions\label{conclusions}}
We have presented detailed numerical calculations of the $\beta$-factor in a PCW. A key step has been to adopt mixed boundary conditions, i.e., active  Dirichlet boundary conditions at the terminations of the waveguide and PMLs at the other boundaries to treat the radiation modes. Based on this approach we calculated the coupling rate from a quantum emitter to different optical channels in a PCW. Our results show that the coupling from the emitter to the radiation continuum is highly suppressed compared to an emitter in homogenous medium. The spatial dependence of $\gamma_{\textrm{rad}}$ quantifies that a suppression factor larger than $10$ is achieved for most regions in the PCW and for all frequencies of the waveguide band. As a direct consequence, the $\beta$-factor is close to unity for essentially all emitter locations in the PCW even for moderately-slow light propagation. The detailed simulations confirm the remarkable robustness of the PCW platform against spatial and spectral inhomogeneities and consequently also fabrication imperfections.
Such a high coupling efficiency is of importance for a wide range of photonic quantum technology applications including on-demand single-photon sources, multi-qubit gates\cite{Mahmoodian2016PRL}, and single-photon transistors\cite{Chang2007NPHYS,Witthaut2012EL}.

\section*{Acknowledgments}

We would like to thank Yuntian Chen for fruitful discussion regarding the finite-element simulations. We gratefully acknowledge financial support from the European Research Council (ERC Advanced Grant "SCALE"), Innovation Fund Denmark (Quantum Innovation Center "Qubiz"), and the Danish Council for Independent Research.

\appendix
\section*{Appendix}
\subsection{Influence of the simulation parameters on $\gamma_\textrm{rad}$ \label{ap:Convergance_tests}}

\begin{figure}[h!]
  \includegraphics[width=86mm]{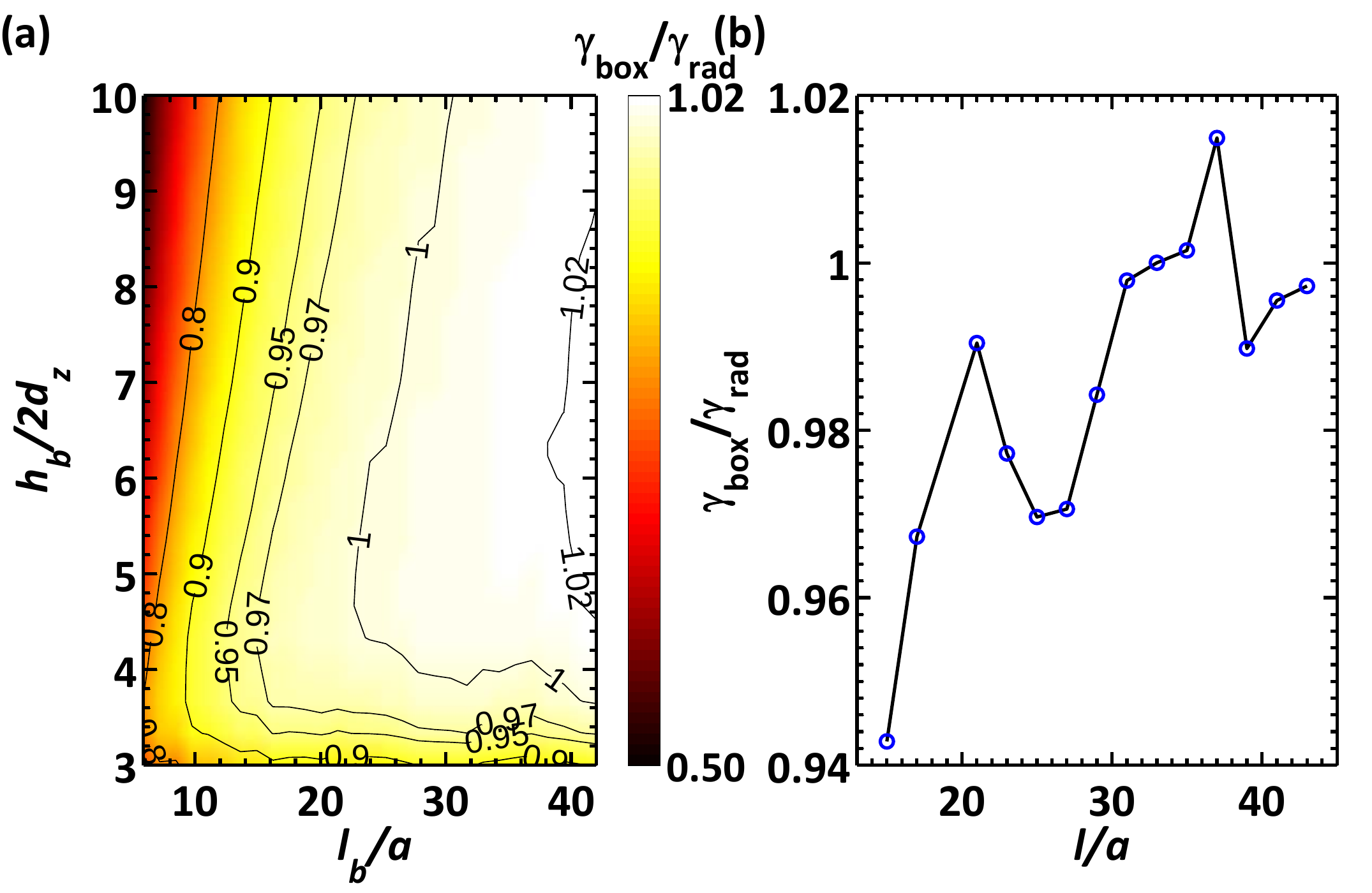}\\
  \caption{Dependence of $\gamma_\textrm{box}$ on the size of the integration box. For $l_b/a>25$ and $h_b/2d_z>4$, $\gamma_\textrm{box}$ fluctuates by less than $5\%$. (b) Dependence of $\gamma_\textrm{box}$ on the actual size of the simulation domain. Both (a) and (b) are calculated for $y$-oriented dipoles in the anti-node of the $E_y$ field. The frequency of the dipole corresponds to $n_g=58$.}\label{conv_test}
\end{figure}

Figure \ref{conv_test} presents some of the convergence tests carried out to ensure the validity of the simulations and to justify the choice of the radiation box size. We choose $\gamma_\textrm{box}/\gamma_\textrm{rad}$ as the target parameter for the convergence tests, where $\gamma_\textrm{box}$ is the amount of radiation captured by the radiation box of size $h_b$ and $l_b$, and $\gamma_\textrm{rad}$ is the value reported in Fig. \ref{spaitial_map_gammarad}.  From Fig. \ref{conv_test}a, we conclude that for $l_b/a>25$ and $h_b/2d_z>4$ the value of $\gamma_\textrm{box}$ is independent of the size of the box to within $5\%$. Furthermore, the convergence of $\gamma_\textrm{rad}$ with the size of the actual simulation domain is plotted in Fig. \ref{conv_test}(b) displaying a similar precision. These convergence tests were carried out for a $y$-oriented dipole at the $E_y$-antinode and with $n_g=58$. We repeated the same tests for dipoles at a few more positions, orientations, and frequencies with very similar results.

\subsection{Position and frequency dependence of coupling to radiation modes in a photonic crystal.\label{ap:PhC_radiationmodes}}

\begin{figure}[t!]
  \includegraphics[width=86mm]{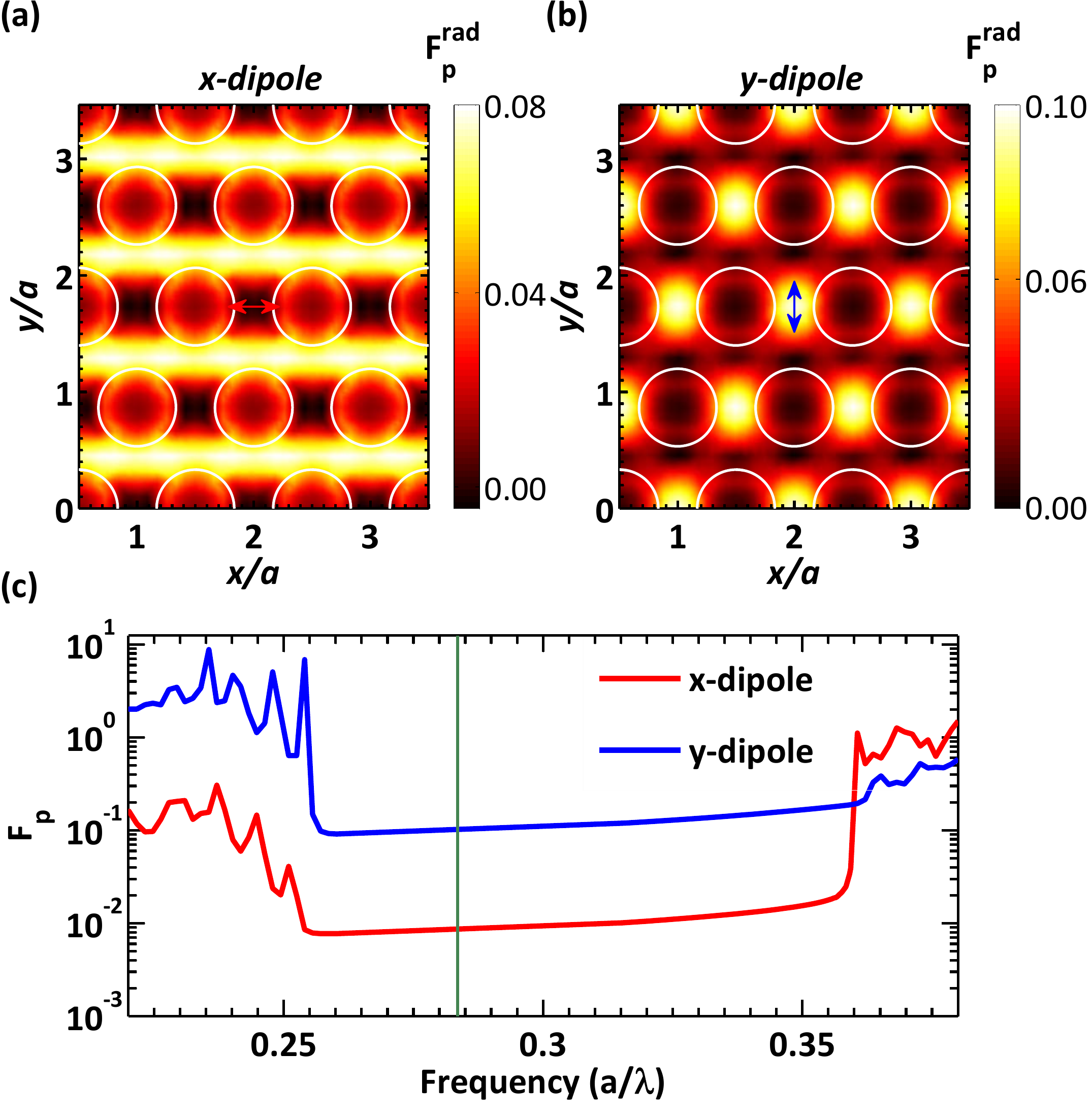}\\
  \caption{(a) and (b) The spatial dependence of $F_p^\textrm{rad}$ for $x$- and $y$-oriented dipoles inside the bandgap of a photonic crystal membrane. The maximum of $F_p^\textrm{rad}$ is $\sim 1/10$ and its minimum value is $\sim 1/168$. The frequency of the emitter corresponds to the vertical green line in part (c). (c) The frequency dependence of $F_p^\textrm{rad}$ for the two dipole positions and orientations shown in parts (a) and (b). }\label{phc_gammarad}
\end{figure}

As a comparison, we present the position and frequency dependence of $F_p^\textrm{rad}$ for dipoles located in a photonic-crystal membrane without any waveguide defect region. These simulations were carried out in a similar fashion as for the PCW case, however they did not require active boundary conditions as the photonic crystal already suppresses the light propagation and hence PML boundary conditions are adequate.  Figure \ref{phc_gammarad} maps out the position dependence of $F_p^\textrm{rad}$ inside the bandgap of a photonic crystal for two orthogonal dipole orientations. Furthermore, the frequency dependence of $F_p^{\textrm{rad}}$ for an emitter in the photonic crystal is displayed. The bandgap of the photonic crystal extends from $a/\lambda=0.256$ to $a/\lambda=0.360$. The main feature is the inhibition of spontaneous emission inside the bandgap of the photonic crystal, which reaches values as high as $168$. These values are very similar to what is found in  PCWs (see Fig. \ref{spaitial_map_gammarad}), and hence we conclude that the missing row of holes in the PCW does not significantly alter the coupling to the radiation modes. We mention that these results compare very well to the values reported in \cite{Koenderink2006JOSAB}.

\bibliography{bigbib}

\begin{thebibliography}{46}%
\makeatletter
\providecommand \@ifxundefined [1]{%
 \@ifx{#1\undefined}
}%
\providecommand \@ifnum [1]{%
 \ifnum #1\expandafter \@firstoftwo
 \else \expandafter \@secondoftwo
 \fi
}%
\providecommand \@ifx [1]{%
 \ifx #1\expandafter \@firstoftwo
 \else \expandafter \@secondoftwo
 \fi
}%
\providecommand \natexlab [1]{#1}%
\providecommand \enquote  [1]{``#1''}%
\providecommand \bibnamefont  [1]{#1}%
\providecommand \bibfnamefont [1]{#1}%
\providecommand \citenamefont [1]{#1}%
\providecommand \href@noop [0]{\@secondoftwo}%
\providecommand \href [0]{\begingroup \@sanitize@url \@href}%
\providecommand \@href[1]{\@@startlink{#1}\@@href}%
\providecommand \@@href[1]{\endgroup#1\@@endlink}%
\providecommand \@sanitize@url [0]{\catcode `\\12\catcode `\$12\catcode
  `\&12\catcode `\#12\catcode `\^12\catcode `\_12\catcode `\%12\relax}%
\providecommand \@@startlink[1]{}%
\providecommand \@@endlink[0]{}%
\providecommand \url  [0]{\begingroup\@sanitize@url \@url }%
\providecommand \@url [1]{\endgroup\@href {#1}{\urlprefix }}%
\providecommand \urlprefix  [0]{URL }%
\providecommand \Eprint [0]{\href }%
\providecommand \doibase [0]{http://dx.doi.org/}%
\providecommand \selectlanguage [0]{\@gobble}%
\providecommand \bibinfo  [0]{\@secondoftwo}%
\providecommand \bibfield  [0]{\@secondoftwo}%
\providecommand \translation [1]{[#1]}%
\providecommand \BibitemOpen [0]{}%
\providecommand \bibitemStop [0]{}%
\providecommand \bibitemNoStop [0]{.\EOS\space}%
\providecommand \EOS [0]{\spacefactor3000\relax}%
\providecommand \BibitemShut  [1]{\csname bibitem#1\endcsname}%
\let\auto@bib@innerbib\@empty
%</preamble>
\bibitem [{\citenamefont {Purcell}(1946)}]{Purcell1946PR}%
  \BibitemOpen
  \bibfield  {author} {\bibinfo {author} {\bibfnamefont {E.~M.}\ \bibnamefont
  {Purcell}},\ }\href@noop {} {\bibfield  {journal} {\bibinfo  {journal} {Phys.
  Rev.}\ }\textbf {\bibinfo {volume} {69}},\ \bibinfo {pages} {681} (\bibinfo
  {year} {1946})}\BibitemShut {NoStop}%
\bibitem [{\citenamefont {Bykov}(1975)}]{Bykov1975}%
  \BibitemOpen
  \bibfield  {author} {\bibinfo {author} {\bibfnamefont {V.~P.}\ \bibnamefont
  {Bykov}},\ }\href@noop {} {\bibfield  {journal} {\bibinfo  {journal} {Sov. J.
  Quant. Electron.}\ }\textbf {\bibinfo {volume} {4}},\ \bibinfo {pages} {861}
  (\bibinfo {year} {1975})}\BibitemShut {NoStop}%
\bibitem [{\citenamefont {Yablonovitch}(1987)}]{Yablonovitch1987PRL}%
  \BibitemOpen
  \bibfield  {author} {\bibinfo {author} {\bibfnamefont {E.}~\bibnamefont
  {Yablonovitch}},\ }\href@noop {} {\bibfield  {journal} {\bibinfo  {journal}
  {Phys. Rev. Lett.}\ }\textbf {\bibinfo {volume} {58}},\ \bibinfo {pages}
  {2059} (\bibinfo {year} {1987})}\BibitemShut {NoStop}%
\bibitem [{\citenamefont {Lodahl}\ \emph {et~al.}(2004)\citenamefont {Lodahl},
  \citenamefont {van Driel}, \citenamefont {Nikolaev}, \citenamefont {Irman},
  \citenamefont {Overgaag}, \citenamefont {Vanmaekelbergh},\ and\ \citenamefont
  {Vos}}]{Lodahl2004Nature}%
  \BibitemOpen
  \bibfield  {author} {\bibinfo {author} {\bibfnamefont {P.}~\bibnamefont
  {Lodahl}}, \bibinfo {author} {\bibfnamefont {A.~F.}\ \bibnamefont {van
  Driel}}, \bibinfo {author} {\bibfnamefont {I.~S.}\ \bibnamefont {Nikolaev}},
  \bibinfo {author} {\bibfnamefont {A.}~\bibnamefont {Irman}}, \bibinfo
  {author} {\bibfnamefont {K.}~\bibnamefont {Overgaag}}, \bibinfo {author}
  {\bibfnamefont {D.}~\bibnamefont {Vanmaekelbergh}}, \ and\ \bibinfo {author}
  {\bibfnamefont {W.~L.}\ \bibnamefont {Vos}},\ }\href@noop {} {\bibfield
  {journal} {\bibinfo  {journal} {Nature}\ }\textbf {\bibinfo {volume} {430}},\
  \bibinfo {pages} {654} (\bibinfo {year} {2004})}\BibitemShut {NoStop}%
\bibitem [{\citenamefont {G\'erard}\ \emph {et~al.}(1998)\citenamefont
  {G\'erard}, \citenamefont {Sermage}, \citenamefont {Gayral}, \citenamefont
  {Legrand}, \citenamefont {Costard},\ and\ \citenamefont
  {Thierry-Mieg}}]{Gerard1998PRL}%
  \BibitemOpen
  \bibfield  {author} {\bibinfo {author} {\bibfnamefont {J.~M.}\ \bibnamefont
  {G\'erard}}, \bibinfo {author} {\bibfnamefont {B.}~\bibnamefont {Sermage}},
  \bibinfo {author} {\bibfnamefont {B.}~\bibnamefont {Gayral}}, \bibinfo
  {author} {\bibfnamefont {B.}~\bibnamefont {Legrand}}, \bibinfo {author}
  {\bibfnamefont {E.}~\bibnamefont {Costard}}, \ and\ \bibinfo {author}
  {\bibfnamefont {V.}~\bibnamefont {Thierry-Mieg}},\ }\href@noop {} {\bibfield
  {journal} {\bibinfo  {journal} {Phys. Rev. Lett.}\ }\textbf {\bibinfo
  {volume} {81}} (\bibinfo {year} {1998})}\BibitemShut {NoStop}%
\bibitem [{\citenamefont {Lund-Hansen}\ \emph {et~al.}(2008)\citenamefont
  {Lund-Hansen}, \citenamefont {Stobbe}, \citenamefont {Julsgaard},
  \citenamefont {Thyrrestrup}, \citenamefont {S\"unner}, \citenamefont {Kamp},
  \citenamefont {Forchel},\ and\ \citenamefont {Lodahl}}]{Lund-Hansen2008PRL}%
  \BibitemOpen
  \bibfield  {author} {\bibinfo {author} {\bibfnamefont {T.}~\bibnamefont
  {Lund-Hansen}}, \bibinfo {author} {\bibfnamefont {S.}~\bibnamefont {Stobbe}},
  \bibinfo {author} {\bibfnamefont {B.}~\bibnamefont {Julsgaard}}, \bibinfo
  {author} {\bibfnamefont {H.}~\bibnamefont {Thyrrestrup}}, \bibinfo {author}
  {\bibfnamefont {T.}~\bibnamefont {S\"unner}}, \bibinfo {author}
  {\bibfnamefont {M.}~\bibnamefont {Kamp}}, \bibinfo {author} {\bibfnamefont
  {A.}~\bibnamefont {Forchel}}, \ and\ \bibinfo {author} {\bibfnamefont
  {P.}~\bibnamefont {Lodahl}},\ }\href@noop {} {\bibfield  {journal} {\bibinfo
  {journal} {Phys. Rev. Lett.}\ }\textbf {\bibinfo {volume} {101}},\ \bibinfo
  {pages} {113903} (\bibinfo {year} {2008})}\BibitemShut {NoStop}%
\bibitem [{\citenamefont {Akimov}\ \emph {et~al.}(2007)\citenamefont {Akimov},
  \citenamefont {Mukherjee}, \citenamefont {Yu}, \citenamefont {Chang},
  \citenamefont {Zibrov}, \citenamefont {Hemmer}, \citenamefont {Park},\ and\
  \citenamefont {Lukin}}]{Akimov2007Nature}%
  \BibitemOpen
  \bibfield  {author} {\bibinfo {author} {\bibfnamefont {A.~V.}\ \bibnamefont
  {Akimov}}, \bibinfo {author} {\bibfnamefont {A.}~\bibnamefont {Mukherjee}},
  \bibinfo {author} {\bibfnamefont {C.~L.}\ \bibnamefont {Yu}}, \bibinfo
  {author} {\bibfnamefont {D.~E.}\ \bibnamefont {Chang}}, \bibinfo {author}
  {\bibfnamefont {A.~S.}\ \bibnamefont {Zibrov}}, \bibinfo {author}
  {\bibfnamefont {P.~R.}\ \bibnamefont {Hemmer}}, \bibinfo {author}
  {\bibfnamefont {H.}~\bibnamefont {Park}}, \ and\ \bibinfo {author}
  {\bibfnamefont {M.~D.}\ \bibnamefont {Lukin}},\ }\href@noop {} {\bibfield
  {journal} {\bibinfo  {journal} {Nature}\ }\textbf {\bibinfo {volume} {450}},\
  \bibinfo {pages} {402} (\bibinfo {year} {2007})}\BibitemShut {NoStop}%
\bibitem [{\citenamefont {Wang}\ \emph {et~al.}(2011)\citenamefont {Wang},
  \citenamefont {Stobbe},\ and\ \citenamefont {Lodahl}}]{Wang2011PRL}%
  \BibitemOpen
  \bibfield  {author} {\bibinfo {author} {\bibfnamefont {Q.}~\bibnamefont
  {Wang}}, \bibinfo {author} {\bibfnamefont {S.}~\bibnamefont {Stobbe}}, \ and\
  \bibinfo {author} {\bibfnamefont {P.}~\bibnamefont {Lodahl}},\ }\href@noop {}
  {\bibfield  {journal} {\bibinfo  {journal} {Phys. Rev. Lett.}\ }\textbf
  {\bibinfo {volume} {107}},\ \bibinfo {pages} {167404} (\bibinfo {year}
  {2011})}\BibitemShut {NoStop}%
\bibitem [{\citenamefont {Lodahl}\ \emph {et~al.}(2015)\citenamefont {Lodahl},
  \citenamefont {Mahmoodian},\ and\ \citenamefont {Stobbe}}]{Lodahl2015RMP}%
  \BibitemOpen
  \bibfield  {author} {\bibinfo {author} {\bibfnamefont {P.}~\bibnamefont
  {Lodahl}}, \bibinfo {author} {\bibfnamefont {S.}~\bibnamefont {Mahmoodian}},
  \ and\ \bibinfo {author} {\bibfnamefont {S.}~\bibnamefont {Stobbe}},\
  }\href@noop {} {\bibfield  {journal} {\bibinfo  {journal} {Rev. Mod. Phys.}\
  }\textbf {\bibinfo {volume} {87}},\ \bibinfo {pages} {347} (\bibinfo {year}
  {2015})}\BibitemShut {NoStop}%
\bibitem [{\citenamefont {Javadi}\ \emph {et~al.}(2015)\citenamefont {Javadi},
  \citenamefont {S{\"o}llner}, \citenamefont {Arcari}, \citenamefont {Hansen},
  \citenamefont {Midolo}, \citenamefont {Mahmoodian}, \citenamefont
  {Kir{\v{s}}ansk{\.e}}, \citenamefont {Pregnolato}, \citenamefont {Lee},
  \citenamefont {Song}, \citenamefont {Stobbe},\ and\ \citenamefont
  {Lodahl}}]{Javadi2015NCOM}%
  \BibitemOpen
  \bibfield  {author} {\bibinfo {author} {\bibfnamefont {A.}~\bibnamefont
  {Javadi}}, \bibinfo {author} {\bibfnamefont {I.}~\bibnamefont {S{\"o}llner}},
  \bibinfo {author} {\bibfnamefont {M.}~\bibnamefont {Arcari}}, \bibinfo
  {author} {\bibfnamefont {S.~L.}\ \bibnamefont {Hansen}}, \bibinfo {author}
  {\bibfnamefont {L.}~\bibnamefont {Midolo}}, \bibinfo {author} {\bibfnamefont
  {S.}~\bibnamefont {Mahmoodian}}, \bibinfo {author} {\bibfnamefont
  {G.}~\bibnamefont {Kir{\v{s}}ansk{\.e}}}, \bibinfo {author} {\bibfnamefont
  {T.}~\bibnamefont {Pregnolato}}, \bibinfo {author} {\bibfnamefont
  {E.}~\bibnamefont {Lee}}, \bibinfo {author} {\bibfnamefont {J.}~\bibnamefont
  {Song}}, \bibinfo {author} {\bibfnamefont {S.}~\bibnamefont {Stobbe}}, \ and\
  \bibinfo {author} {\bibfnamefont {P.}~\bibnamefont {Lodahl}},\ }\href@noop {}
  {\bibfield  {journal} {\bibinfo  {journal} {Nat. Commun.}\ }\textbf {\bibinfo
  {volume} {6}},\ \bibinfo {pages} {8655} (\bibinfo {year} {2015})}\BibitemShut
  {NoStop}%
\bibitem [{\citenamefont {Pinotsi}\ \emph {et~al.}(2011)\citenamefont
  {Pinotsi}, \citenamefont {Fallahi}, \citenamefont {Miguel-Sanchez},\ and\
  \citenamefont {Imamoglu}}]{Pinotsi2001IEEEJQP}%
  \BibitemOpen
  \bibfield  {author} {\bibinfo {author} {\bibfnamefont {D.}~\bibnamefont
  {Pinotsi}}, \bibinfo {author} {\bibfnamefont {P.}~\bibnamefont {Fallahi}},
  \bibinfo {author} {\bibfnamefont {J.}~\bibnamefont {Miguel-Sanchez}}, \ and\
  \bibinfo {author} {\bibfnamefont {A.}~\bibnamefont {Imamoglu}},\ }\href@noop
  {} {\bibfield  {journal} {\bibinfo  {journal} {IEEE J. Quant Electron.}\
  }\textbf {\bibinfo {volume} {47}},\ \bibinfo {pages} {1371} (\bibinfo {year}
  {2011})}\BibitemShut {NoStop}%
\bibitem [{\citenamefont {Carter}\ \emph {et~al.}(2013)\citenamefont {Carter},
  \citenamefont {Sweeney}, \citenamefont {Kim}, \citenamefont {Kim},
  \citenamefont {Solenov}, \citenamefont {Economou}, \citenamefont {Reinecke},
  \citenamefont {Yang}, \citenamefont {Bracker},\ and\ \citenamefont
  {Gammon}}]{Carter2013NPHOT}%
  \BibitemOpen
  \bibfield  {author} {\bibinfo {author} {\bibfnamefont {S.~G.}\ \bibnamefont
  {Carter}}, \bibinfo {author} {\bibfnamefont {T.~M.}\ \bibnamefont {Sweeney}},
  \bibinfo {author} {\bibfnamefont {M.}~\bibnamefont {Kim}}, \bibinfo {author}
  {\bibfnamefont {C.~S.}\ \bibnamefont {Kim}}, \bibinfo {author} {\bibfnamefont
  {D.}~\bibnamefont {Solenov}}, \bibinfo {author} {\bibfnamefont {S.~E.}\
  \bibnamefont {Economou}}, \bibinfo {author} {\bibfnamefont {T.~L.}\
  \bibnamefont {Reinecke}}, \bibinfo {author} {\bibfnamefont {L.}~\bibnamefont
  {Yang}}, \bibinfo {author} {\bibfnamefont {A.~S.}\ \bibnamefont {Bracker}}, \
  and\ \bibinfo {author} {\bibfnamefont {D.}~\bibnamefont {Gammon}},\
  }\href@noop {} {\bibfield  {journal} {\bibinfo  {journal} {Nat. Photonics}\
  }\textbf {\bibinfo {volume} {7}},\ \bibinfo {pages} {329} (\bibinfo {year}
  {2013})}\BibitemShut {NoStop}%
\bibitem [{\citenamefont {Makhonin}\ \emph {et~al.}(2014)\citenamefont
  {Makhonin}, \citenamefont {Dixon}, \citenamefont {Coles}, \citenamefont
  {Royall}, \citenamefont {Luxmoore}, \citenamefont {Clarke}, \citenamefont
  {Hugues}, \citenamefont {Skolnick},\ and\ \citenamefont
  {Fox}}]{makhonin2014waveguide}%
  \BibitemOpen
  \bibfield  {author} {\bibinfo {author} {\bibfnamefont {M.~N.}\ \bibnamefont
  {Makhonin}}, \bibinfo {author} {\bibfnamefont {J.~E.}\ \bibnamefont {Dixon}},
  \bibinfo {author} {\bibfnamefont {R.~J.}\ \bibnamefont {Coles}}, \bibinfo
  {author} {\bibfnamefont {B.}~\bibnamefont {Royall}}, \bibinfo {author}
  {\bibfnamefont {I.~J.}\ \bibnamefont {Luxmoore}}, \bibinfo {author}
  {\bibfnamefont {E.}~\bibnamefont {Clarke}}, \bibinfo {author} {\bibfnamefont
  {M.}~\bibnamefont {Hugues}}, \bibinfo {author} {\bibfnamefont {M.~S.}\
  \bibnamefont {Skolnick}}, \ and\ \bibinfo {author} {\bibfnamefont {A.~M.}\
  \bibnamefont {Fox}},\ }\href {\doibase 10.1021/nl5032937} {\bibfield
  {journal} {\bibinfo  {journal} {Nano lett.}\ }\textbf {\bibinfo {volume}
  {14}},\ \bibinfo {pages} {6997} (\bibinfo {year} {2014})}\BibitemShut
  {NoStop}%
\bibitem [{\citenamefont {Lon\v{c}ar}\ and\ \citenamefont
  {Faraon}(2013)}]{Loncar2013MRS}%
  \BibitemOpen
  \bibfield  {author} {\bibinfo {author} {\bibfnamefont {M.}~\bibnamefont
  {Lon\v{c}ar}}\ and\ \bibinfo {author} {\bibfnamefont {A.}~\bibnamefont
  {Faraon}},\ }\href@noop {} {\bibfield  {journal} {\bibinfo  {journal} {MRS
  Bull.}\ }\textbf {\bibinfo {volume} {38}},\ \bibinfo {pages} {144} (\bibinfo
  {year} {2013})}\BibitemShut {NoStop}%
\bibitem [{\citenamefont {Sipahigil}\ \emph {et~al.}(2016)\citenamefont
  {Sipahigil}, \citenamefont {Evans}, \citenamefont {Sukachev}, \citenamefont
  {Burek}, \citenamefont {Borregaard}, \citenamefont {Bhaskar}, \citenamefont
  {Nguyen}, \citenamefont {Pacheco}, \citenamefont {Atikian}, \citenamefont
  {Meuwly}, \citenamefont {Camacho}, \citenamefont {Jelezko}, \citenamefont
  {Bielejec}, \citenamefont {Park}, \citenamefont {Lon{\v c}ar},\ and\
  \citenamefont {Lukin}}]{Sipahigil2016Science}%
  \BibitemOpen
  \bibfield  {author} {\bibinfo {author} {\bibfnamefont {A.}~\bibnamefont
  {Sipahigil}}, \bibinfo {author} {\bibfnamefont {R.~E.}\ \bibnamefont
  {Evans}}, \bibinfo {author} {\bibfnamefont {D.~D.}\ \bibnamefont {Sukachev}},
  \bibinfo {author} {\bibfnamefont {M.~J.}\ \bibnamefont {Burek}}, \bibinfo
  {author} {\bibfnamefont {J.}~\bibnamefont {Borregaard}}, \bibinfo {author}
  {\bibfnamefont {M.~K.}\ \bibnamefont {Bhaskar}}, \bibinfo {author}
  {\bibfnamefont {C.~T.}\ \bibnamefont {Nguyen}}, \bibinfo {author}
  {\bibfnamefont {J.~L.}\ \bibnamefont {Pacheco}}, \bibinfo {author}
  {\bibfnamefont {H.~A.}\ \bibnamefont {Atikian}}, \bibinfo {author}
  {\bibfnamefont {C.}~\bibnamefont {Meuwly}}, \bibinfo {author} {\bibfnamefont
  {R.~M.}\ \bibnamefont {Camacho}}, \bibinfo {author} {\bibfnamefont
  {F.}~\bibnamefont {Jelezko}}, \bibinfo {author} {\bibfnamefont
  {E.}~\bibnamefont {Bielejec}}, \bibinfo {author} {\bibfnamefont
  {H.}~\bibnamefont {Park}}, \bibinfo {author} {\bibfnamefont {M.}~\bibnamefont
  {Lon{\v c}ar}}, \ and\ \bibinfo {author} {\bibfnamefont {M.~D.}\ \bibnamefont
  {Lukin}},\ }\href {\doibase 10.1126/science.aah6875} {\bibfield  {journal}
  {\bibinfo  {journal} {Science}\ }\textbf {\bibinfo {volume} {354}},\ \bibinfo
  {pages} {847} (\bibinfo {year} {2016})}\BibitemShut {NoStop}%
\bibitem [{\citenamefont {Tiecke}\ \emph {et~al.}(2014)\citenamefont {Tiecke},
  \citenamefont {Thompson}, \citenamefont {de~Leon}, \citenamefont {Liu},
  \citenamefont {Vuleti{\'c}},\ and\ \citenamefont {Lukin}}]{Tiecke2014Nature}%
  \BibitemOpen
  \bibfield  {author} {\bibinfo {author} {\bibfnamefont {T.}~\bibnamefont
  {Tiecke}}, \bibinfo {author} {\bibfnamefont {J.}~\bibnamefont {Thompson}},
  \bibinfo {author} {\bibfnamefont {N.}~\bibnamefont {de~Leon}}, \bibinfo
  {author} {\bibfnamefont {L.}~\bibnamefont {Liu}}, \bibinfo {author}
  {\bibfnamefont {V.}~\bibnamefont {Vuleti{\'c}}}, \ and\ \bibinfo {author}
  {\bibfnamefont {M.}~\bibnamefont {Lukin}},\ }\href@noop {} {\bibfield
  {journal} {\bibinfo  {journal} {Nature}\ }\textbf {\bibinfo {volume} {508}},\
  \bibinfo {pages} {241} (\bibinfo {year} {2014})}\BibitemShut {NoStop}%
\bibitem [{\citenamefont {Goban}\ \emph {et~al.}(2014)\citenamefont {Goban},
  \citenamefont {Hung}, \citenamefont {Yu}, \citenamefont {Hood}, \citenamefont
  {Muniz}, \citenamefont {Lee}, \citenamefont {Martin.}, \citenamefont
  {McClung}, \citenamefont {Choi}, \citenamefont {Chang}, \citenamefont
  {Painter},\ and\ \citenamefont {Kimble}}]{Goban2014Ncom}%
  \BibitemOpen
  \bibfield  {author} {\bibinfo {author} {\bibfnamefont {A.}~\bibnamefont
  {Goban}}, \bibinfo {author} {\bibfnamefont {C.-L.}\ \bibnamefont {Hung}},
  \bibinfo {author} {\bibfnamefont {S.-P.}\ \bibnamefont {Yu}}, \bibinfo
  {author} {\bibfnamefont {J.~D.}\ \bibnamefont {Hood}}, \bibinfo {author}
  {\bibfnamefont {J.~A.}\ \bibnamefont {Muniz}}, \bibinfo {author}
  {\bibfnamefont {J.~H.}\ \bibnamefont {Lee}}, \bibinfo {author} {\bibfnamefont
  {M.~J.}\ \bibnamefont {Martin.}}, \bibinfo {author} {\bibfnamefont {A.~C.}\
  \bibnamefont {McClung}}, \bibinfo {author} {\bibfnamefont {K.~S.}\
  \bibnamefont {Choi}}, \bibinfo {author} {\bibfnamefont {D.~E.}\ \bibnamefont
  {Chang}}, \bibinfo {author} {\bibfnamefont {O.}~\bibnamefont {Painter}}, \
  and\ \bibinfo {author} {\bibfnamefont {H.~J.}\ \bibnamefont {Kimble}},\
  }\href@noop {} {\bibfield  {journal} {\bibinfo  {journal} {Nat. Commun.}\
  }\textbf {\bibinfo {volume} {5}},\ \bibinfo {pages} {3808} (\bibinfo {year}
  {2014})}\BibitemShut {NoStop}%
\bibitem [{\citenamefont {Koenderink}\ \emph {et~al.}(2006)\citenamefont
  {Koenderink}, \citenamefont {Kafesaki}, \citenamefont {Soukoulis},\ and\
  \citenamefont {Sandoghdar}}]{Koenderink2006JOSAB}%
  \BibitemOpen
  \bibfield  {author} {\bibinfo {author} {\bibfnamefont {A.~F.}\ \bibnamefont
  {Koenderink}}, \bibinfo {author} {\bibfnamefont {M.}~\bibnamefont
  {Kafesaki}}, \bibinfo {author} {\bibfnamefont {C.~M.}\ \bibnamefont
  {Soukoulis}}, \ and\ \bibinfo {author} {\bibfnamefont {V.}~\bibnamefont
  {Sandoghdar}},\ }\href@noop {} {\bibfield  {journal} {\bibinfo  {journal} {J.
  Opt. Soc. Am. B.}\ }\textbf {\bibinfo {volume} {23}},\ \bibinfo {pages}
  {1196} (\bibinfo {year} {2006})}\BibitemShut {NoStop}%
\bibitem [{\citenamefont {\protect{Manga Rao}}\ and\ \citenamefont
  {Hughes}(2007{\natexlab{a}})}]{MangaRao2007PRL}%
  \BibitemOpen
  \bibfield  {author} {\bibinfo {author} {\bibfnamefont {V.~S.~C.}\
  \bibnamefont {\protect{Manga Rao}}}\ and\ \bibinfo {author} {\bibfnamefont
  {S.}~\bibnamefont {Hughes}},\ }\href@noop {} {\bibfield  {journal} {\bibinfo
  {journal} {Phys. Rev. Lett.}\ }\textbf {\bibinfo {volume} {99}},\ \bibinfo
  {pages} {193901} (\bibinfo {year} {2007}{\natexlab{a}})}\BibitemShut
  {NoStop}%
\bibitem [{\citenamefont {Lecamp}\ \emph {et~al.}(2007)\citenamefont {Lecamp},
  \citenamefont {Lalanne},\ and\ \citenamefont {Hugonin}}]{LeCamp2007PRL}%
  \BibitemOpen
  \bibfield  {author} {\bibinfo {author} {\bibfnamefont {G.}~\bibnamefont
  {Lecamp}}, \bibinfo {author} {\bibfnamefont {P.}~\bibnamefont {Lalanne}}, \
  and\ \bibinfo {author} {\bibfnamefont {J.~P.}\ \bibnamefont {Hugonin}},\
  }\href@noop {} {\bibfield  {journal} {\bibinfo  {journal} {Phys. Rev. Lett.}\
  }\textbf {\bibinfo {volume} {99}},\ \bibinfo {pages} {023902} (\bibinfo
  {year} {2007})}\BibitemShut {NoStop}%
\bibitem [{\citenamefont {Dewhurst}\ \emph {et~al.}(2010)\citenamefont
  {Dewhurst}, \citenamefont {Granados}, \citenamefont {Ellis}, \citenamefont
  {Bennett}, \citenamefont {Patel}, \citenamefont {Farrer}, \citenamefont
  {Anderson}, \citenamefont {Jones}, \citenamefont {Ritchie},\ and\
  \citenamefont {Shields}}]{Dewhurst2010APL}%
  \BibitemOpen
  \bibfield  {author} {\bibinfo {author} {\bibfnamefont {S.~J.}\ \bibnamefont
  {Dewhurst}}, \bibinfo {author} {\bibfnamefont {D.}~\bibnamefont {Granados}},
  \bibinfo {author} {\bibfnamefont {D.~J.~P.}\ \bibnamefont {Ellis}}, \bibinfo
  {author} {\bibfnamefont {A.~J.}\ \bibnamefont {Bennett}}, \bibinfo {author}
  {\bibfnamefont {R.~B.}\ \bibnamefont {Patel}}, \bibinfo {author}
  {\bibfnamefont {I.}~\bibnamefont {Farrer}}, \bibinfo {author} {\bibfnamefont
  {D.}~\bibnamefont {Anderson}}, \bibinfo {author} {\bibfnamefont {G.~A.~C.}\
  \bibnamefont {Jones}}, \bibinfo {author} {\bibfnamefont {D.~A.}\ \bibnamefont
  {Ritchie}}, \ and\ \bibinfo {author} {\bibfnamefont {A.~J.}\ \bibnamefont
  {Shields}},\ }\href@noop {} {\bibfield  {journal} {\bibinfo  {journal} {Appl.
  Phys. Lett.}\ }\textbf {\bibinfo {volume} {96}},\ \bibinfo {pages} {031109}
  (\bibinfo {year} {2010})}\BibitemShut {NoStop}%
\bibitem [{\citenamefont {Hoang}\ \emph {et~al.}(2012)\citenamefont {Hoang},
  \citenamefont {Beetz}, \citenamefont {Midolo}, \citenamefont {Skacel},
  \citenamefont {Lermer}, \citenamefont {Kamp}, \citenamefont {H\"ofling},
  \citenamefont {Balet}, \citenamefont {Chauvin},\ and\ \citenamefont
  {Fiore}}]{Hoang2012APL}%
  \BibitemOpen
  \bibfield  {author} {\bibinfo {author} {\bibfnamefont {T.~B.}\ \bibnamefont
  {Hoang}}, \bibinfo {author} {\bibfnamefont {J.}~\bibnamefont {Beetz}},
  \bibinfo {author} {\bibfnamefont {L.}~\bibnamefont {Midolo}}, \bibinfo
  {author} {\bibfnamefont {M.}~\bibnamefont {Skacel}}, \bibinfo {author}
  {\bibfnamefont {M.}~\bibnamefont {Lermer}}, \bibinfo {author} {\bibfnamefont
  {M.}~\bibnamefont {Kamp}}, \bibinfo {author} {\bibfnamefont {S.}~\bibnamefont
  {H\"ofling}}, \bibinfo {author} {\bibfnamefont {L.}~\bibnamefont {Balet}},
  \bibinfo {author} {\bibfnamefont {N.}~\bibnamefont {Chauvin}}, \ and\
  \bibinfo {author} {\bibfnamefont {A.}~\bibnamefont {Fiore}},\ }\href@noop {}
  {\bibfield  {journal} {\bibinfo  {journal} {Appl. Phys. Lett.}\ }\textbf
  {\bibinfo {volume} {100}},\ \bibinfo {pages} {061122} (\bibinfo {year}
  {2012})}\BibitemShut {NoStop}%
\bibitem [{\citenamefont {Laucht}\ \emph {et~al.}(2012)\citenamefont {Laucht},
  \citenamefont {P\"{u}tz}, \citenamefont {G\"{u}nthner}, \citenamefont
  {Hauke}, \citenamefont {Saive}, \citenamefont {Fr\'{e}d\'{e}rick},
  \citenamefont {Bichler}, \citenamefont {Amann}, \citenamefont {Holleitner},
  \citenamefont {Kaniber},\ and\ \citenamefont {Finley}}]{Laucht2012PRX}%
  \BibitemOpen
  \bibfield  {author} {\bibinfo {author} {\bibfnamefont {A.}~\bibnamefont
  {Laucht}}, \bibinfo {author} {\bibfnamefont {S.}~\bibnamefont {P\"{u}tz}},
  \bibinfo {author} {\bibfnamefont {T.}~\bibnamefont {G\"{u}nthner}}, \bibinfo
  {author} {\bibfnamefont {N.}~\bibnamefont {Hauke}}, \bibinfo {author}
  {\bibfnamefont {R.}~\bibnamefont {Saive}}, \bibinfo {author} {\bibfnamefont
  {S.}~\bibnamefont {Fr\'{e}d\'{e}rick}}, \bibinfo {author} {\bibfnamefont
  {M.}~\bibnamefont {Bichler}}, \bibinfo {author} {\bibfnamefont {M.-C.}\
  \bibnamefont {Amann}}, \bibinfo {author} {\bibfnamefont {A.}~\bibnamefont
  {Holleitner}}, \bibinfo {author} {\bibfnamefont {M.}~\bibnamefont {Kaniber}},
  \ and\ \bibinfo {author} {\bibfnamefont {J.~J.}\ \bibnamefont {Finley}},\
  }\href@noop {} {\bibfield  {journal} {\bibinfo  {journal} {Phys. Rev. X}\
  }\textbf {\bibinfo {volume} {2}},\ \bibinfo {pages} {011014} (\bibinfo {year}
  {2012})}\BibitemShut {NoStop}%
\bibitem [{\citenamefont {Arcari}\ \emph {et~al.}(2014)\citenamefont {Arcari},
  \citenamefont {S\"ollner}, \citenamefont {Javadi}, \citenamefont
  {Lindskov~Hansen}, \citenamefont {Mahmoodian}, \citenamefont {Liu},
  \citenamefont {Thyrrestrup}, \citenamefont {Lee}, \citenamefont {Song},
  \citenamefont {Stobbe},\ and\ \citenamefont {Lodahl}}]{Arcari2014PRL}%
  \BibitemOpen
  \bibfield  {author} {\bibinfo {author} {\bibfnamefont {M.}~\bibnamefont
  {Arcari}}, \bibinfo {author} {\bibfnamefont {I.}~\bibnamefont {S\"ollner}},
  \bibinfo {author} {\bibfnamefont {A.}~\bibnamefont {Javadi}}, \bibinfo
  {author} {\bibfnamefont {S.}~\bibnamefont {Lindskov~Hansen}}, \bibinfo
  {author} {\bibfnamefont {S.}~\bibnamefont {Mahmoodian}}, \bibinfo {author}
  {\bibfnamefont {J.}~\bibnamefont {Liu}}, \bibinfo {author} {\bibfnamefont
  {H.}~\bibnamefont {Thyrrestrup}}, \bibinfo {author} {\bibfnamefont {E.~H.}\
  \bibnamefont {Lee}}, \bibinfo {author} {\bibfnamefont {J.~D.}\ \bibnamefont
  {Song}}, \bibinfo {author} {\bibfnamefont {S.}~\bibnamefont {Stobbe}}, \ and\
  \bibinfo {author} {\bibfnamefont {P.}~\bibnamefont {Lodahl}},\ }\href
  {\doibase 10.1103/PhysRevLett.113.093603} {\bibfield  {journal} {\bibinfo
  {journal} {Phys. Rev. Lett.}\ }\textbf {\bibinfo {volume} {113}},\ \bibinfo
  {pages} {093603} (\bibinfo {year} {2014})}\BibitemShut {NoStop}%
\bibitem [{\citenamefont {Joannopoulos}\ \emph {et~al.}(2008)\citenamefont
  {Joannopoulos}, \citenamefont {Johnson}, \citenamefont {Winn},\ and\
  \citenamefont {Meade}}]{JoannopoulosBook}%
  \BibitemOpen
  \bibfield  {author} {\bibinfo {author} {\bibfnamefont {J.~D.}\ \bibnamefont
  {Joannopoulos}}, \bibinfo {author} {\bibfnamefont {S.~G.}\ \bibnamefont
  {Johnson}}, \bibinfo {author} {\bibfnamefont {J.~N.}\ \bibnamefont {Winn}}, \
  and\ \bibinfo {author} {\bibfnamefont {R.~D.}\ \bibnamefont {Meade}},\
  }\href@noop {} {\emph {\bibinfo {title} {Photonic Crystals: Molding the Flow
  of Light}}}\ (\bibinfo  {publisher} {Princeton University Press},\ \bibinfo
  {year} {2008})\BibitemShut {NoStop}%
\bibitem [{\citenamefont {Johnson}\ and\ \citenamefont
  {Joannopoulos}(2001)}]{Johnson2001OE}%
  \BibitemOpen
  \bibfield  {author} {\bibinfo {author} {\bibfnamefont {S.~G.}\ \bibnamefont
  {Johnson}}\ and\ \bibinfo {author} {\bibfnamefont {J.~D.}\ \bibnamefont
  {Joannopoulos}},\ }\href@noop {} {\bibfield  {journal} {\bibinfo  {journal}
  {Opt. Express}\ }\textbf {\bibinfo {volume} {8}},\ \bibinfo {pages} {173}
  (\bibinfo {year} {2001})}\BibitemShut {NoStop}%
\bibitem [{\citenamefont {\protect{Manga Rao}}\ and\ \citenamefont
  {Hughes}(2007{\natexlab{b}})}]{MangaRao2007PRB}%
  \BibitemOpen
  \bibfield  {author} {\bibinfo {author} {\bibfnamefont {V.~S.~C.}\
  \bibnamefont {\protect{Manga Rao}}}\ and\ \bibinfo {author} {\bibfnamefont
  {S.}~\bibnamefont {Hughes}},\ }\href@noop {} {\bibfield  {journal} {\bibinfo
  {journal} {Phys. Rev. B}\ }\textbf {\bibinfo {volume} {75}},\ \bibinfo
  {pages} {205437} (\bibinfo {year} {2007}{\natexlab{b}})}\BibitemShut
  {NoStop}%
\bibitem [{\citenamefont {Oskooi}\ \emph {et~al.}(2008)\citenamefont {Oskooi},
  \citenamefont {Zhang}, \citenamefont {Avniel},\ and\ \citenamefont
  {Johnson}}]{Oskooi2008OE}%
  \BibitemOpen
  \bibfield  {author} {\bibinfo {author} {\bibfnamefont {A.~F.}\ \bibnamefont
  {Oskooi}}, \bibinfo {author} {\bibfnamefont {L.}~\bibnamefont {Zhang}},
  \bibinfo {author} {\bibfnamefont {Y.}~\bibnamefont {Avniel}}, \ and\ \bibinfo
  {author} {\bibfnamefont {S.~G.}\ \bibnamefont {Johnson}},\ }\href@noop {}
  {\bibfield  {journal} {\bibinfo  {journal} {Opt. Express}\ }\textbf {\bibinfo
  {volume} {16}},\ \bibinfo {pages} {11376} (\bibinfo {year}
  {2008})}\BibitemShut {NoStop}%
\bibitem [{\citenamefont {Hughes}\ \emph {et~al.}(2005)\citenamefont {Hughes},
  \citenamefont {Ramunno}, \citenamefont {Young},\ and\ \citenamefont
  {Sipe}}]{Hughes2005PRL}%
  \BibitemOpen
  \bibfield  {author} {\bibinfo {author} {\bibfnamefont {S.}~\bibnamefont
  {Hughes}}, \bibinfo {author} {\bibfnamefont {L.}~\bibnamefont {Ramunno}},
  \bibinfo {author} {\bibfnamefont {J.~F.}\ \bibnamefont {Young}}, \ and\
  \bibinfo {author} {\bibfnamefont {J.~E.}\ \bibnamefont {Sipe}},\ }\href@noop
  {} {\bibfield  {journal} {\bibinfo  {journal} {Phys. Rev. Lett.}\ }\textbf
  {\bibinfo {volume} {94}},\ \bibinfo {pages} {033903} (\bibinfo {year}
  {2005})}\BibitemShut {NoStop}%
\bibitem [{\citenamefont {Mazoyer}\ \emph {et~al.}(2009)\citenamefont
  {Mazoyer}, \citenamefont {Hugonin},\ and\ \citenamefont
  {Lalanne}}]{Mazoyer2009PRL}%
  \BibitemOpen
  \bibfield  {author} {\bibinfo {author} {\bibfnamefont {S.}~\bibnamefont
  {Mazoyer}}, \bibinfo {author} {\bibfnamefont {J.~P.}\ \bibnamefont
  {Hugonin}}, \ and\ \bibinfo {author} {\bibfnamefont {P.}~\bibnamefont
  {Lalanne}},\ }\href@noop {} {\bibfield  {journal} {\bibinfo  {journal} {Phys.
  Rev. Lett.}\ }\textbf {\bibinfo {volume} {103}},\ \bibinfo {pages} {063903}
  (\bibinfo {year} {2009})}\BibitemShut {NoStop}%
\bibitem [{\citenamefont {Savona}(2011)}]{savona2011PRB}%
  \BibitemOpen
  \bibfield  {author} {\bibinfo {author} {\bibfnamefont {V.}~\bibnamefont
  {Savona}},\ }\href@noop {} {\bibfield  {journal} {\bibinfo  {journal} {Phys.
  Rev. B}\ }\textbf {\bibinfo {volume} {83}},\ \bibinfo {pages} {085301}
  (\bibinfo {year} {2011})}\BibitemShut {NoStop}%
\bibitem [{\citenamefont {Smolka}\ \emph {et~al.}(2011)\citenamefont {Smolka},
  \citenamefont {Thyrrestrup}, \citenamefont {Sapienza}, \citenamefont
  {Lehmann}, \citenamefont {Rix}, \citenamefont {Froufe-P\'erez}, \citenamefont
  {Garc\'ia},\ and\ \citenamefont {Lodahl}}]{Smolka2011NJP}%
  \BibitemOpen
  \bibfield  {author} {\bibinfo {author} {\bibfnamefont {S.}~\bibnamefont
  {Smolka}}, \bibinfo {author} {\bibfnamefont {H.}~\bibnamefont {Thyrrestrup}},
  \bibinfo {author} {\bibfnamefont {L.}~\bibnamefont {Sapienza}}, \bibinfo
  {author} {\bibfnamefont {T.~B.}\ \bibnamefont {Lehmann}}, \bibinfo {author}
  {\bibfnamefont {K.~R.}\ \bibnamefont {Rix}}, \bibinfo {author} {\bibfnamefont
  {L.~S.}\ \bibnamefont {Froufe-P\'erez}}, \bibinfo {author} {\bibfnamefont
  {P.~D.}\ \bibnamefont {Garc\'ia}}, \ and\ \bibinfo {author} {\bibfnamefont
  {P.}~\bibnamefont {Lodahl}},\ }\href@noop {} {\bibfield  {journal} {\bibinfo
  {journal} {New J. Phys.}\ }\textbf {\bibinfo {volume} {13}},\ \bibinfo
  {pages} {063044} (\bibinfo {year} {2011})}\BibitemShut {NoStop}%
\bibitem [{\citenamefont {Novotny}\ and\ \citenamefont
  {Hecht}(2007)}]{NanoOpticsBook}%
  \BibitemOpen
  \bibfield  {author} {\bibinfo {author} {\bibfnamefont {L.}~\bibnamefont
  {Novotny}}\ and\ \bibinfo {author} {\bibfnamefont {B.}~\bibnamefont
  {Hecht}},\ }\href@noop {} {\emph {\bibinfo {title} {Principles of
  nano-optics}}}\ (\bibinfo  {publisher} {Cambridge University Press},\
  \bibinfo {year} {2007})\BibitemShut {NoStop}%
\bibitem [{\citenamefont {de~Lasson}\ \emph {et~al.}(2012)\citenamefont
  {de~Lasson}, \citenamefont {Christensen}, \citenamefont {M{\o}rk},\ and\
  \citenamefont {Gregersen}}]{De2012JOSA}%
  \BibitemOpen
  \bibfield  {author} {\bibinfo {author} {\bibfnamefont {J.~R.}\ \bibnamefont
  {de~Lasson}}, \bibinfo {author} {\bibfnamefont {T.}~\bibnamefont
  {Christensen}}, \bibinfo {author} {\bibfnamefont {J.}~\bibnamefont
  {M{\o}rk}}, \ and\ \bibinfo {author} {\bibfnamefont {N.}~\bibnamefont
  {Gregersen}},\ }\href@noop {} {\bibfield  {journal} {\bibinfo  {journal} {J.
  Opt. Soc. Am. A.}\ }\textbf {\bibinfo {volume} {29}},\ \bibinfo {pages}
  {1237} (\bibinfo {year} {2012})}\BibitemShut {NoStop}%
\bibitem [{\citenamefont {Chen}\ \emph {et~al.}(2010)\citenamefont {Chen},
  \citenamefont {Nielsen}, \citenamefont {Gregersen}, \citenamefont {Lodahl},\
  and\ \citenamefont {M{\o}rk}}]{Chen2010PRB}%
  \BibitemOpen
  \bibfield  {author} {\bibinfo {author} {\bibfnamefont {Y.}~\bibnamefont
  {Chen}}, \bibinfo {author} {\bibfnamefont {T.~R.}\ \bibnamefont {Nielsen}},
  \bibinfo {author} {\bibfnamefont {N.}~\bibnamefont {Gregersen}}, \bibinfo
  {author} {\bibfnamefont {P.}~\bibnamefont {Lodahl}}, \ and\ \bibinfo {author}
  {\bibfnamefont {J.}~\bibnamefont {M{\o}rk}},\ }\href@noop {} {\bibfield
  {journal} {\bibinfo  {journal} {Phys. Rev. B}\ }\textbf {\bibinfo {volume}
  {81}},\ \bibinfo {pages} {125431} (\bibinfo {year} {2010})}\BibitemShut
  {NoStop}%
\bibitem [{\citenamefont {Berenger}(1994)}]{berenger1994perfectly}%
  \BibitemOpen
  \bibfield  {author} {\bibinfo {author} {\bibfnamefont {J.-P.}\ \bibnamefont
  {Berenger}},\ }\href@noop {} {\bibfield  {journal} {\bibinfo  {journal} {J.
  Comp. Phys.}\ }\textbf {\bibinfo {volume} {114}},\ \bibinfo {pages} {185}
  (\bibinfo {year} {1994})}\BibitemShut {NoStop}%
\bibitem [{\citenamefont {Chew}\ and\ \citenamefont
  {Weedon}(1994)}]{chew19943MOTL}%
  \BibitemOpen
  \bibfield  {author} {\bibinfo {author} {\bibfnamefont {W.~C.}\ \bibnamefont
  {Chew}}\ and\ \bibinfo {author} {\bibfnamefont {W.~H.}\ \bibnamefont
  {Weedon}},\ }\href@noop {} {\bibfield  {journal} {\bibinfo  {journal}
  {Microwave and optical technology letters}\ }\textbf {\bibinfo {volume}
  {7}},\ \bibinfo {pages} {599} (\bibinfo {year} {1994})}\BibitemShut {NoStop}%
\bibitem [{\citenamefont {Wasley}\ \emph {et~al.}(2012)\citenamefont {Wasley},
  \citenamefont {Luxmoore}, \citenamefont {Coles}, \citenamefont {Clarke},
  \citenamefont {Fox},\ and\ \citenamefont {Skolnick}}]{Wasley2012APL}%
  \BibitemOpen
  \bibfield  {author} {\bibinfo {author} {\bibfnamefont {N.~A.}\ \bibnamefont
  {Wasley}}, \bibinfo {author} {\bibfnamefont {I.}~\bibnamefont {Luxmoore}},
  \bibinfo {author} {\bibfnamefont {R.}~\bibnamefont {Coles}}, \bibinfo
  {author} {\bibfnamefont {E.}~\bibnamefont {Clarke}}, \bibinfo {author}
  {\bibfnamefont {A.}~\bibnamefont {Fox}}, \ and\ \bibinfo {author}
  {\bibfnamefont {M.}~\bibnamefont {Skolnick}},\ }\href@noop {} {\bibfield
  {journal} {\bibinfo  {journal} {Appl. Phys. Lett.}\ }\textbf {\bibinfo
  {volume} {101}},\ \bibinfo {pages} {051116} (\bibinfo {year}
  {2012})}\BibitemShut {NoStop}%
\bibitem [{\citenamefont {Javadi}\ \emph {et~al.}(2014)\citenamefont {Javadi},
  \citenamefont {Maibom}, \citenamefont {Sapienza}, \citenamefont
  {Thyrrestrup}, \citenamefont {Garc\'{i}a},\ and\ \citenamefont
  {Lodahl}}]{Javadi2014OE}%
  \BibitemOpen
  \bibfield  {author} {\bibinfo {author} {\bibfnamefont {A.}~\bibnamefont
  {Javadi}}, \bibinfo {author} {\bibfnamefont {S.}~\bibnamefont {Maibom}},
  \bibinfo {author} {\bibfnamefont {L.}~\bibnamefont {Sapienza}}, \bibinfo
  {author} {\bibfnamefont {H.}~\bibnamefont {Thyrrestrup}}, \bibinfo {author}
  {\bibfnamefont {P.~D.}\ \bibnamefont {Garc\'{i}a}}, \ and\ \bibinfo {author}
  {\bibfnamefont {P.}~\bibnamefont {Lodahl}},\ }\href {\doibase
  10.1364/OE.22.030992} {\bibfield  {journal} {\bibinfo  {journal} {Opt.
  Express}\ }\textbf {\bibinfo {volume} {22}},\ \bibinfo {pages} {30992}
  (\bibinfo {year} {2014})}\BibitemShut {NoStop}%
\bibitem [{\citenamefont {John}\ and\ \citenamefont
  {Wang}(1991)}]{John1991PRB}%
  \BibitemOpen
  \bibfield  {author} {\bibinfo {author} {\bibfnamefont {S.}~\bibnamefont
  {John}}\ and\ \bibinfo {author} {\bibfnamefont {J.}~\bibnamefont {Wang}},\
  }\href@noop {} {\bibfield  {journal} {\bibinfo  {journal} {Phys. Rev. B}\
  }\textbf {\bibinfo {volume} {43}},\ \bibinfo {pages} {12772} (\bibinfo {year}
  {1991})}\BibitemShut {NoStop}%
\bibitem [{\citenamefont {Douglas}\ \emph {et~al.}(2015)\citenamefont
  {Douglas}, \citenamefont {Habibian}, \citenamefont {Hung}, \citenamefont
  {Gorshkov}, \citenamefont {Kimble},\ and\ \citenamefont
  {Chang}}]{Douglas2015NPHOT}%
  \BibitemOpen
  \bibfield  {author} {\bibinfo {author} {\bibfnamefont {J.~S.}\ \bibnamefont
  {Douglas}}, \bibinfo {author} {\bibfnamefont {H.}~\bibnamefont {Habibian}},
  \bibinfo {author} {\bibfnamefont {C.-L.}\ \bibnamefont {Hung}}, \bibinfo
  {author} {\bibfnamefont {A.}~\bibnamefont {Gorshkov}}, \bibinfo {author}
  {\bibfnamefont {H.~J.}\ \bibnamefont {Kimble}}, \ and\ \bibinfo {author}
  {\bibfnamefont {D.~E.}\ \bibnamefont {Chang}},\ }\href@noop {} {\bibfield
  {journal} {\bibinfo  {journal} {Nat. Photonics}\ }\textbf {\bibinfo {volume}
  {9}},\ \bibinfo {pages} {326} (\bibinfo {year} {2015})}\BibitemShut {NoStop}%
\bibitem [{\citenamefont {Munro}\ \emph {et~al.}(2016)\citenamefont {Munro},
  \citenamefont {Kwek},\ and\ \citenamefont {Chang}}]{Munro2016Arxiv}%
  \BibitemOpen
  \bibfield  {author} {\bibinfo {author} {\bibfnamefont {E.}~\bibnamefont
  {Munro}}, \bibinfo {author} {\bibfnamefont {L.~C.}\ \bibnamefont {Kwek}}, \
  and\ \bibinfo {author} {\bibfnamefont {D.~E.}\ \bibnamefont {Chang}},\
  }\href@noop {} {\bibfield  {journal} {\bibinfo  {journal} {arXiv:1604.02893}\
  } (\bibinfo {year} {2016})}\BibitemShut {NoStop}%
\bibitem [{\citenamefont {Calaj{\'o}}\ \emph {et~al.}(2016)\citenamefont
  {Calaj{\'o}}, \citenamefont {Ciccarello}, \citenamefont {Chang},\ and\
  \citenamefont {Rabl}}]{Calajo2016PRA}%
  \BibitemOpen
  \bibfield  {author} {\bibinfo {author} {\bibfnamefont {G.}~\bibnamefont
  {Calaj{\'o}}}, \bibinfo {author} {\bibfnamefont {F.}~\bibnamefont
  {Ciccarello}}, \bibinfo {author} {\bibfnamefont {D.}~\bibnamefont {Chang}}, \
  and\ \bibinfo {author} {\bibfnamefont {P.}~\bibnamefont {Rabl}},\ }\href@noop
  {} {\bibfield  {journal} {\bibinfo  {journal} {Phys. Rev. A}\ }\textbf
  {\bibinfo {volume} {93}},\ \bibinfo {pages} {033833} (\bibinfo {year}
  {2016})}\BibitemShut {NoStop}%
\bibitem [{\citenamefont {Mahmoodian}\ \emph {et~al.}(2016)\citenamefont
  {Mahmoodian}, \citenamefont {Lodahl},\ and\ \citenamefont
  {S\o{}rensen}}]{Mahmoodian2016PRL}%
  \BibitemOpen
  \bibfield  {author} {\bibinfo {author} {\bibfnamefont {S.}~\bibnamefont
  {Mahmoodian}}, \bibinfo {author} {\bibfnamefont {P.}~\bibnamefont {Lodahl}},
  \ and\ \bibinfo {author} {\bibfnamefont {A.~S.}\ \bibnamefont
  {S\o{}rensen}},\ }\href {\doibase 10.1103/PhysRevLett.117.240501} {\bibfield
  {journal} {\bibinfo  {journal} {Phys. Rev. Lett.}\ }\textbf {\bibinfo
  {volume} {117}},\ \bibinfo {pages} {240501} (\bibinfo {year}
  {2016})}\BibitemShut {NoStop}%
\bibitem [{\citenamefont {Chang}\ \emph {et~al.}(2007)\citenamefont {Chang},
  \citenamefont {S{\o}rensen}, \citenamefont {Demler},\ and\ \citenamefont
  {Lukin}}]{Chang2007NPHYS}%
  \BibitemOpen
  \bibfield  {author} {\bibinfo {author} {\bibfnamefont {D.~E.}\ \bibnamefont
  {Chang}}, \bibinfo {author} {\bibfnamefont {A.~S.}\ \bibnamefont
  {S{\o}rensen}}, \bibinfo {author} {\bibfnamefont {E.~A.}\ \bibnamefont
  {Demler}}, \ and\ \bibinfo {author} {\bibfnamefont {M.~D.}\ \bibnamefont
  {Lukin}},\ }\href@noop {} {\bibfield  {journal} {\bibinfo  {journal} {Nat.
  Phys.}\ }\textbf {\bibinfo {volume} {3}},\ \bibinfo {pages} {807} (\bibinfo
  {year} {2007})}\BibitemShut {NoStop}%
\bibitem [{\citenamefont {Witthaut}\ \emph {et~al.}(2012)\citenamefont
  {Witthaut}, \citenamefont {Lukin},\ and\ \citenamefont
  {S{\o}rensen}}]{Witthaut2012EL}%
  \BibitemOpen
  \bibfield  {author} {\bibinfo {author} {\bibfnamefont {D.}~\bibnamefont
  {Witthaut}}, \bibinfo {author} {\bibfnamefont {M.~D.}\ \bibnamefont {Lukin}},
  \ and\ \bibinfo {author} {\bibfnamefont {A.~S.}\ \bibnamefont
  {S{\o}rensen}},\ }\href@noop {} {\bibfield  {journal} {\bibinfo  {journal}
  {Europhys. Lett.}\ }\textbf {\bibinfo {volume} {97}},\ \bibinfo {pages}
  {50007} (\bibinfo {year} {2012})}\BibitemShut {NoStop}%
\end{thebibliography}%

\end{document}